\documentclass [12pt,reqno]{elsarticle}
\usepackage{amssymb,amsmath,amsthm}
\usepackage{graphicx}
\usepackage[T1]{fontenc}
\usepackage{tikz}
\usepackage{float}
\usepackage{eurosym}
\usepackage{subcaption}
\usepackage{verbatim}
\usepackage[refpage]{nomencl}
\usepackage{natbib}
\usepackage{relsize}
\usepackage{hyperref}
\usepackage[a4paper]{geometry}

\newtheorem{theorem}{Theorem}[section]

\newtheorem{remark}[theorem]{Remark}

\theoremstyle{definition}
\newtheorem{definition}[theorem]{Definition}
\newtheorem{example}[theorem]{Example}
\newtheorem*{notation}{Notation}

\newtheorem*{properties}{Properties}

\newcommand{\abs}[1]{\left\vert#1\right\vert}
\newcommand{\R}{\ensuremath{\mathbb{R}}}

\renewcommand{\sp}{\par\vspace{1mm}}

\usepackage{comment}
\usepackage{tikz}
\usetikzlibrary{matrix}

\usepackage{natbib}
\usepackage{geometry}
\usepackage{fleqn}
\usepackage{graphicx}
\usepackage{txfonts}

\makeatletter
\def\Links{\tagsleft@true}\def\Rechts{\tagsleft@false}
\makeatother

\begin{document}
\date{\today}

\begin{frontmatter}
\title{New definitions (measures) of skewness, mean and dispersion of fuzzy numbers.\\
- by way of a new representation as parameterized curves.\tnoteref{t1}}
\tnotetext[t1]{This work was supported by the National Science Center (Poland), under Grant 394311, 2017/27/B/HS4/01881 “Selected methods supporting project management, taking into consideration various stakeholder groups and using type-2 fuzzy numbers”. }
\author{Jan Schneider}
\ead{jan.schneider@pwr.edu.pl}
\address{Faculty of Computer Science and Management,
Wrocław University of Science and Technology,
ul. Ignacego Łukasiewicza 5, 50-371 Wrocław, Poland}

\begin{abstract}
We give a geometrically motivated measure of skewness, define a mean value triangle number, and dispersion (in that order) of a fuzzy number without reference or seeking analogy to the namesake but parallel concepts in probability theory. These measures come about by way of a new representation of fuzzy numbers as parameterized curves respectively their associated tangent bundle.
Importantly skewness and dispersion are given as functions of $\alpha$ (the degree of membership) and such may be given separately and pointwise at each $\alpha$-level, as well as overall.
This allows for e.g., when a mathematical model is formulated in fuzzy numbers, to run optimization programs level-wise thereby encapsuling with deliberate accuracy the involved membership functions' characteristics while increasing the computational complexity by only a multiplicative factor compared to the same program formulated in real variables and parameters. \par
As an example the work offers a contribution to the recently very popular fuzzy mean-variance-skewness portfolio optimization.
\end{abstract}

\begin{keyword}
fuzzy numbers \sep parametric curves \sep polar coordinates \sep skewness of a fuzzy number \sep dispersion of a fuzzy number \sep portfolio optimization
\end{keyword}
\end{frontmatter}

\tableofcontents

\section{Introduction}

\subsection{Motivation}
In a natural language the term ``skewness'' is in itself a vague concept, but when it occurs in conversation is well understood intuitively by speakers of the English language as ``deviating from symmetry'', `` deviating from a straight line'', ``slant to the left or the right'', obviously without mathematically specifying its meaning.\sp

In scientific context the term is most often associated on the grounds of probability theory and statistics where many mathematical concepts have been introduced to capture and fit the intuitively understood meaning of the term with respect to probability distributions.

This evolution of the concept started with 1985 Pearson~\cite{Pearson1895} who gives his mode and moment coefficients of skewness of a probability distribution, and today quite a number of measures of probabilistic skewness exist in parallel, many more are being developed, each useful for different given contexts.\sp

In search of universal properties which a skewness measure (coefficient) $\gamma$ should satisfy to be considered useful in the conceptual structure of random variables van Zwet in 1964~\cite{Zwe64} named four desiderata, namely

\begin{enumerate}
\item[\textbf{P.1}] A scale or location change for a random variable does not alter $\gamma.$ Thus, if $Y=cX+d$ for
$c>O$ and $-\infty<d<\infty$ then $\gamma(X)=\gamma(Y),$
\item[\textbf{P.2}] For a symmetric distribution $\gamma = 0,$
\item[\textbf{P.3}] If $Y=-X$ then $\gamma(Y)=-\gamma(X),$
\item[\textbf{P.4}] If $F$ and $G$ CDFs for $X$ and $Y$ as above and $F<G$ then $\gamma(X)<\gamma(Y).$
\end{enumerate}

Although in this article much attention is devoted to underline the fact that the theory of probability and that of fuzzy sets provide for two  completely different ways of evaluating and expressing uncertainty, we see that the fuzzy analogues of P.1. to P.3. should hold for a fuzzy measure of skewness, whereas it is debatable, whether P.4 can be coherently transferred. This depends e.g. on whether the linguistic value being modeled by a fuzzy number has an underlying base variable or not. This matter should be discussed separately, and is not part of the discussion in this paper. \sp

This article attempts to give a definition of skewness in the context of membership functions of fuzzy numbers, a definition which best matches the intuition naturally partnered with the term, when practitioners use it in the context of uncertainty best described and adequately quantifiable as fuzzy information by fuzzy numbers. \sp

Although we develop our own definition of skewness independently, and it arises naturally from the perspective of looking at fuzzy numbers as parameterized curves which own their specific differential geometry, it is practically impossible to discuss the matter without referencing to the various measures grounded and employed in probability and statistics, as the hitherto existing measures of fuzzy skewness are built with reference to the latter, that is as quite literal translations. This fact in itself does not necessarily imply a lack of functional capacity of those measures, but as much as one may track the roots of fuzzy set theory as \textit{ensembles flous} to Karl Menger's probabilistic metric spaces~\cite{Men51} the two concepts of quantified uncertainty, probability and fuzziness, have grown apart very much apart into separate domains and the need for an independent definition of skewness that would be rooted in the very own characteristics of fuzzy sets seems to be there. \sp




This next section gives a short recap of concepts of center, variability and shape in the theory of probability and discusses the various analogues which function in the fuzzy literature, but also suggests some hitherto uncovered directions.

\subsection{Concepts of center, variability and shape in the theory of probability}
In probability theory one may distinguish between firstly

\begin{itemize}
\item \textit{parameter free skewness coefficients}, \sp
such as the original quartile (also called Bowley skewness, Galton's measure of skewness) coefficient
\begin{equation}\label{EquationBowleySkewness}
S = \frac{Q_1 - 2\,Q_2 + Q_3}{Q_3-Q_1}
\end{equation}
or the following skewness function introduced by R. A. Groeneveld and G. Meeden in 1984~\cite{GM84}
\begin{equation}\label{EquationGMskewness}
\gamma(u) = \frac{F^{-1}(u) + F^{-1}(1-u) - 2F^{-1}(1/2)}{F^{-1}(u) - F^{-1}(1-u)}.
\end{equation}
which incorporates~\eqref{EquationBowleySkewness} at $u=\dfrac{3}{4}$.

\item \textit{and secondly parameter measures of skewness} \sp

the most prominent of which are Pearson's mode and moment skewness coefficients, which assume as primary the parameters of mode and mean (variance follows from mean) going $Mean \rightarrow Variance \, (SD) \rightarrow Skewness, Kurtosis$:

\textbf{1.} Mode Skewness (Pearson's first skewness coefficient)
\begin{equation}\label{EquationPearsonsMode}
S = \frac{\mu-\nu}{\sigma},
\end{equation}

but when speaking of skewness most people think of \sp
\textbf{2.} Pearson's moment coefficient of skewness around the mean
\begin{equation}\label{EquationPearsonsMoment}
S = E\left[\left(\frac{X-\mu}{\sigma}\right)^3\right].
\end{equation}

\item It should be noted that many more measures of skewness do exist and are in permanent use, especially in the field of financial economics.

\end{itemize}

\subsubsection{Transferring concepts of center variability and shape from probability into the theory of fuzzy sets} \sp

As above we divide between skewness measures which do or do not incorporate other descriptive parameters such as mean and variance: \sp

\paragraph{1. Parameter free skewness measures}
The skewness function~\eqref{EquationGMskewness} may appear particularly interesting and a possible subject of future research from the fuzzy number point of view: -
One may construct a fuzzy number $\xi$ by using the CDFs $F_l$ and (reflected) $F_r$ of two random variables $X_l$ and $X_r$ as the left and right fuzzy sides $l(x)$ and $r(x)$ of $\xi$ and transfer and apply~\eqref{EquationGMskewness} directly from probability distributions to fuzzy numbers generated like that. One may also conversely redefine the left and right endpoints of a fuzzy number as the CDFs of two distributions $F_l$ and (reflected) $F_r$ and set for instance:
\begin{equation}
\gamma^{*}(u) = \frac{\gamma_l(u) + \gamma_r(u)}{2}.
\end{equation}
Interestingly, to the author's knowledge, no effort has been taken to transfer~\eqref{EquationGMskewness}, directly into the realm of fuzzy membership functions. \par
But, as in the next approach described in the next paragraph, while this approach is formally-mathematically viable and painless, the interpretational value is difficult and doubtful, because it demands for the linking of the value of a linguistic variable with the CDF of a probability distribution, which are entirely different concepts, although the density function of a unimodal probability distribution looks similar to many people.\sp

\paragraph{2. Analogues of Pearson's mode and moment coefficient in fuzzy set theory}\label{SubsubsectionMomentProbabilityFuzzy}
The bulk of fuzzy literature relates to Pearson's moment coefficient although~\eqref{EquationPearsonsMode} may be transferred directly e.g. by taking
$$ \text{mode = most probable}\quad \cong \quad\text{middle = $C_1(\xi)$},$$
and any of the available ranking indices for mean value in combination with any metric for absolute average deviation. \sp

In bringing over Pearson's moment coefficient two approaches suggest themselves: \sp

\textbf{1.} When one reads Pearson's original 1895 argumentation for the use of the second and third moments it is of physical nature (center of mass, moments of rotational inertia) and curve fitting and one sees that the reasoning may transfer directly, that is technically without reference to probability distributions: \sp

Starting with the \emph{possibilistic mean value} introduced by Goetschel and Voxman~\cite{GV86} (later generalized in~\cite{CF01}, \cite{FM03}), which in our notation (see~Notation~\eqref{EquationParametricDefinition}) is written
\begin{equation}\label{EquationPossibilisticMean}
\mu_{P} = E_{P}(\xi) = \frac{1}{2}\int_0^1 d(\alpha) + u(\alpha)\,d\alpha
\end{equation}
higher \emph{possibilistic} moments are introduced by
\begin{equation}
\mu^n_{P}(\xi) = \frac{1}{2}\int_0^1 \left(d(\alpha)-E_{P}\right)^n\,d\alpha + \frac{1}{2} \int_0^1 \left(u(\alpha)-E_{P}\right)^n\,d\alpha
\end{equation}
with skewness being the third such defined moment
\begin{equation}
S^*(\xi) = \mu^3_{P}(\xi).
\end{equation} \sp

An approach along these lines, in the setting of possibility theory, was taken by E. Vercher, E. and J.D. Berm\'{u}dez~\cite{VB12,VB13} in the context of fuzzy portfolio optimization. \sp

(The values this skewness coefficient may take infinity as the probabilistic counterpart, which is counter-intuitive in many contexts, and its use may often not be appropriate for this or other reasons.) \sp

\textbf{2.} Other authors build a bridge to probabilistic moments by relating/ associating a given membership function $\xi(x)$ with a probability density function $f_{X}(x)$ by either
\begin{enumerate}
\item scaling $$f_{X}(x) = \xi(x)/\int_{-\infty}^{+\infty}\xi(t)dt$$

\item or by $$f_{X}(x) = \xi(x)\frac{d}{dx}\xi(x) \quad ( = \frac{d}{dx} \xi^2(x)) $$\label{second}
\end{enumerate}
both implying the existence of a meaningful underlying random variable $X$ directly associated with \emph{value} $\mathcal{Y}$ of a linguistic \emph{variable} which is represented by the membership function $\xi_{\mathcal{Y}}$.
\begin{figure}[H]
\centering
\setlength{\unitlength}{3947sp}%
\begingroup\makeatletter\ifx\SetFigFont\undefined%
\gdef\SetFigFont#1#2#3#4#5{%
  \reset@font\fontsize{#1}{#2pt}%
  \fontfamily{#3}\fontseries{#4}\fontshape{#5}%
  \selectfont}%
\fi\endgroup%
\begin{picture}(1353,1386)(511,-1024)
\thicklines
{\color[rgb]{0,0,0}\put(751,239){\vector(-1, 0){  0}}
\put(751,239){\vector( 1, 0){900}}
}%
{\color[rgb]{0,0,0}\put(751,-961){\vector(-1, 0){  0}}
\put(751,-961){\vector( 1, 0){900}}
}%
{\color[rgb]{0,0,0}\put(601, 89){\vector( 0,-1){900}}
}%
{\color[rgb]{0,0,0}\put(1801, 89){\vector( 0,-1){900}}
}%
{\color[rgb]{1,0,0}\multiput(1051,164)(6.00000,6.00000){26}{\makebox(6.6667,10.0000){\small.}}
}%
{\color[rgb]{1,0,0}\multiput(1201,164)(6.00000,6.00000){26}{\makebox(6.6667,10.0000){\small.}}
}%
\put(526,239){\makebox(0,0)[lb]{\smash{{\SetFigFont{12}{14.4}{\familydefault}{\mddefault}{\updefault}{\color[rgb]{0,0,0}$X$}%
}}}}
\put(526,-961){\makebox(0,0)[lb]{\smash{{\SetFigFont{12}{14.4}{\familydefault}{\mddefault}{\updefault}{\color[rgb]{0,0,0}$f_X$}%
}}}}
\put(1726,-961){\makebox(0,0)[lb]{\smash{{\SetFigFont{12}{14.4}{\familydefault}{\mddefault}{\updefault}{\color[rgb]{0,0,0}$\xi_{\mathcal{Y}}$}%
}}}}
\put(1726,239){\makebox(0,0)[lb]{\smash{{\SetFigFont{12}{14.4}{\familydefault}{\mddefault}{\updefault}{\color[rgb]{0,0,0}$\mathcal{Y}$}%
}}}}
\end{picture}%
\caption{A non-commutative diagram of linguistic (fuzzy) and probabilistic notions.}
\end{figure}
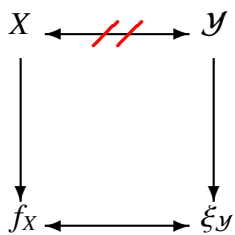

This is at the same time our foremost point of critique against this linking of fuzzy membership functions with probability densities, that this link should be accompanied by an interpretation of the implied link between the represented linguistic value of a linguistic variable, and the random variable, which the density stands for. \par
If such an interpretation cannot be given the identifies given above are mathematically-\-formally correct but without content and not fit for meaningful applications. To give an example: for a trapezoidal fuzzy number (interval) the link $f_{X}(x) = \xi(x)\frac{d}{dx}\xi(x)$ gives a bimodal density function supported outside of $C_1(\xi).$

\begin{remark}
Interestingly the mean value of the probabilistic distribution defined by \textbf{\ref{second}.} coincides with the possibilistic mean value of $\xi$ with weight $\alpha$.
\end{remark}

This approach was taken in e.g.~\cite{LGY15} in the very context of mean-variance-skewness portfolio optimization. \sp


\section{Skewness of triangular (linear) fuzzy numbers}\label{SectionTriangles}
It is assumed that the reader is generally familiar with the concept of fuzzy sets, fuzzy numbers and triangular fuzzy numbers in particular. \par
(For a very clear and exhaustive account see the classic monograph of Buckley~\cite{Buck02}, for a little more recent text with special reference to fuzzy methods in statistics see R. Viertl~\cite{Vie11})\sp

For an understanding of linguistic variables and their linguistic values, which are modeled by fuzzy numbers the reader is best referred to the very founder and main contributor of these concepts in science, i.e. Lotfi A. Zadeh in~\cite{Zad75.1,Zad75.2,Zad75.3,Zad97}. \sp

For easy reference regarding fuzzy numbers and to disambiguate, below are notation and terminology used throughout this particular paper.

\begin{notation}\label{Notation}
In this paper we denote fuzzy numbers by lowercase Greek letters $\xi,\nu,\ldots .$
\noindent By a fuzzy number we understand a fuzzy set on $\R$, which attains $x = 1$ for one and only one $x = m$. We refer to $m$ as the \emph{middle} of $\xi.$ We sometimes shorten ``fuzzy number'' to \emph{FN}.

\noindent A fuzzy number $\xi$ may be given as an ordered function pair of a \emph{left} and a \emph{right side}:
\begin{equation}\label{EquationOrderedPairDefinition}
\xi := \left(l(x),r(x)\right),
\end{equation}
The ordered pair~\eqref{EquationOrderedPairDefinition} may be comprised into a single function $\xi(x)$, which is called the single \emph{membership function} of the fuzzy number $\xi.$ When given as either as a single membership function $\xi(x)$ or as the ordered pair $\left(l(x),r(x)\right)$ we here refer to $\xi$ such given as in \emph{traditional representation}.

The \emph{support} of $l(x)$ is denoted by $[l,m]$, and the support of $r(x)$ by $[m,r]$ and both are onto $[0,1].$ \sp

\noindent We assume invertibility and denote $d(\alpha) := l^{-1}(x)$ and $u(\alpha) := r^{-1}(x)$ and write
\begin{equation}\label{EquationParametricDefinition}
\xi := \left[d(\alpha),u(\alpha)\right]
\end{equation}
for the same fuzzy number $\xi$ in \emph{parametric representation}.
\end{notation} \sp

\begin{figure}[H]
\begin{subfigure}{.45\textwidth}
  \centering
  \includegraphics[height=5cm]{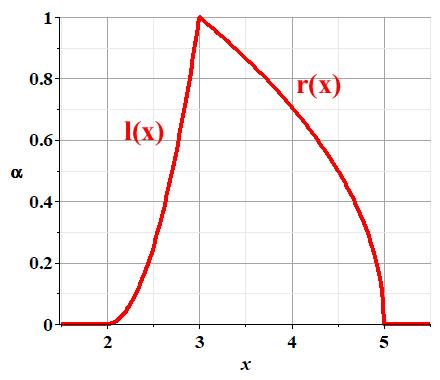}\caption{$\xi$ in traditional representation}
\end{subfigure}
\vspace{0.5cm}
\begin{subfigure}{.45\textwidth}
\centering
  \includegraphics[height=5cm]{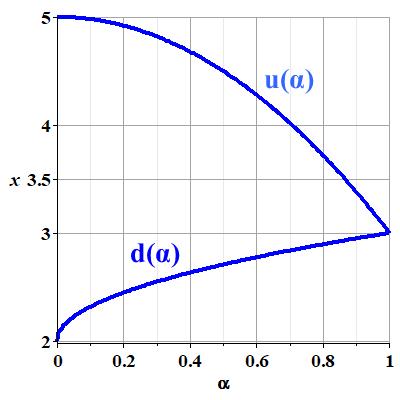}\caption{$\xi$ in parametric representation}
\end{subfigure}
  \caption{A fuzzy number $\xi$ in both traditional and parametric representations}\label{FigureExample1TangentAngle0.jpg}
\end{figure}

For each $\alpha_0\,\in[0,1]$ the closed interval $[d(\alpha_0),u(\alpha_0)]$ is a level set of the membership function $\xi(x)$ and is termed the \emph{$\alpha$-level} or \emph{$\alpha$-cut} of the fuzzy number $\xi$ at $\alpha_0$, denoted by $C_{\alpha_0} (\xi).$  For example $C_1(\xi) = \{m\}$ for fuzzy numbers, as we use the name in this paper.

We denote fuzzy triangular numbers, that is fuzzy numbers with linear sides $l$ and $r$, aka \emph{linear fuzzy numbers} by $tr^*(l,m,r)$ where
\begin{equation}
\xi = tr^*(l,m,r)(x) :=
\begin{cases}
\vspace{2mm}
l(x) = \dfrac{x-l}{m-l} \quad \text{ for } x \in[l,m],\\
r(x) = \dfrac{r-x}{r-m} \quad \text{ for } x \in[m,r].
\end{cases}
\end{equation}
and $l(m) = r(m) = 1,$ and $l,r = 0$ out of $[l,r]$ in traditional representation, \sp

\noindent respectively
\begin{equation}\label{EquationTriangleParametricDefinition}
tr^*[l,m,r](\alpha) = \left[l + \alpha(m-l), r-\alpha(r-m)\right], \quad \alpha\in[0,1]
\end{equation}
in parametric representation.\sp

\bigskip
In this section we will develop an angular skewness coefficient $\gamma_{\xi}$
of a fuzzy triangle number $\xi = tr^*(l,m,r)$ given by:
\begin{equation}\label{EquationGammaTriangle}
\gamma_{\xi}(\alpha) =
\begin{cases}
\pm\:\arccos\,\left(\frac{1}{\sqrt{2}}\cdot\frac{(r-l)}{|\mathbf{r}|}\right), \quad \text{for $|\mathbf{r}|\neq 0$}, \\
0 \qquad \text{ for $m-l = r-m$ (symmetry).}
\end{cases}
\end{equation}
whereby the sign in~\eqref{EquationGammaTriangle} is determined by
\begin{equation}\label{EquationGammaTriangleSign}
 =
\begin{cases}
+  \text{ if }\quad m-l < r-m, \\
-  \text{ if }\quad m-l > r-m,
\end{cases}
\end{equation}
by definition and design taking values
\begin{equation}\label{EquationGammaTriangleValues}
\gamma \in \biggl[- \frac{\pi}{4}, \frac{\pi}{4}\biggr].
\end{equation}

\bigskip
The line of reasoning, which \underline{necessarily} leads straight to this and no other measure $\gamma$ originates from a new perspective of fuzzy numbers as parameterized curves, is laid out below.

\subsection{Derivation of a skewness measure for linear fuzzy numbers}
The line of reasoning is very geometric and graphical, so we show how to arrive at \eqref{EquationGammaTriangle},\eqref{EquationGammaTriangleSign},\eqref{EquationGammaTriangleValues} by looking into a concrete example.\sp

\subsubsection{By example}
\noindent The fuzzy triangle number $tr^*(2,4,5)$ may be written in traditional representation as
\begin{equation}\label{ExampleOneChar}
tr^*(2,4,5)(x) =
\begin{cases}
(x-2)/2 \text{ for } x \in [2,4),\\
1 \text{ for } x=4,\\
(5-x) \text{ for } x = \in (4,5], \\
0 \text{ else}.
\end{cases}
\end{equation}

\noindent or given equivalently in by its nested sequence of $\alpha$-cuts:
\begin{equation}\label{ExampleParam}
C_{\alpha}(tr^*(2,4,5)) = [2+2\alpha, 5 - \alpha],\,\alpha\in[0,1]
\end{equation}
which corresponds to the parametric representation (see \eqref{EquationParametricDefinition}) of $tr^*(2,4,5).$ \sp

Here is the key observation, key to newness, which is central to this paper:\sp

For each $\alpha:$ the $\alpha$-cut given by $[2+2\alpha, 5 - \alpha]$ may be naturally identified with the corresponding point/ row vector  $(2+2\alpha, 5 - \alpha)$ or equivalently column vector $\binom{2+2\alpha}{5-\alpha}$ in the closed half plane $\{x\leq y\} \subseteq \mathbf{R}^2$. \footnote{Thank You, Franck Barthe of \textit{Institut de Math\'{e}matiques de Toulouse}, for this one.}

This equivalence of is shown beneath in \textit{Figure~\ref{FigureExample1Point_3,45_}} for the arbitrarily chosen $\alpha$-level $\alpha = 0.5$:
\begin{figure}[H]
  \centering
  \includegraphics[width=8cm]{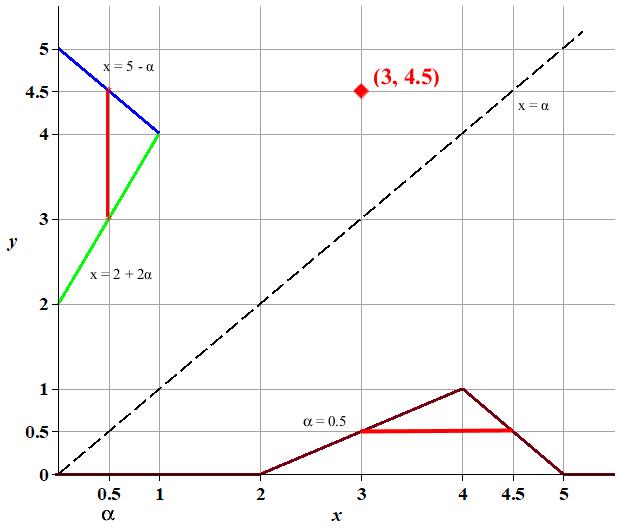}
  \caption{Equivalent objects: \textbf{1.} the level set ($\alpha$-cut) $C_{\frac{1}{2}}\bigl(tr^*(2,4,5)\bigr) = [3,\,4.5]$, \textbf{2.} the interval value of $tr^*[2,4,5](\alpha)$ at $\alpha = \frac{1}{2}$ is $[3,\,4.5],$ and \textbf{3.} the point $\bigl(3,\,4.5\bigr)$ belonging to the half plane $\{x\leq y\}\subseteq\mathbf{R}^2.$}\label{FigureExample1Point_3,45_}
\end{figure}

The identification shown above for the single $\alpha$-value $\alpha = 0.5$ can be of course be performed for every $\alpha\in[0,1]$:\sp

\begin{remark}
By identifying each closed interval $[2+2\alpha, 5 - \alpha] \subset\mathbf{R}$ with the corresponding point $(2+2\alpha, 5 - \alpha) \in \mathbf{R}^2$ one may then define a parameterized curve $\sigma(\alpha)= ( \sigma_1(\alpha),\sigma_2(\alpha), \sigma_3(\alpha) )$ in $\mathbf{R}^3$ given by
\begin{equation}\label{EquationTriangleExample3d}
\sigma(\alpha) = (2+2\alpha, 5 - \alpha,\alpha)
\end{equation}
as shown in Fig.~\ref{FigureTriangleExample3d} below:

\begin{figure}[H]
  \centering
  \includegraphics[width=5cm]{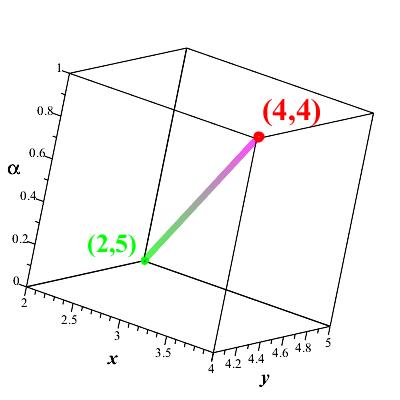}
  \caption{The parameterized curve $\sigma(\alpha) = (2+2\alpha, 5 - \alpha,\alpha)$ corresponding to the fuzzy triangle number $tr^*(2,4,5)$.}
  \label{FigureTriangleExample3d}
\end{figure}

It is clear, that in the case of fuzzy triangle numbers, because of the linearity of all three coordinate functions of $\sigma(\alpha)$ and the resulting similarity of section triangles along the $\{x = y\}$ - line the third dimension may be omitted (as the information, $\alpha$, is given by the first two coordinates), and one may substitute~\eqref{EquationTriangleExample3d} with its projection onto the $xy$ - plane: $\sigma(\alpha)= (2+2\alpha, 5 - \alpha),\, \alpha\in [0,1],$ where $\sigma(0) = (2,5), \sigma(1) = (4,4)$ and for all $\alpha\in[0,1]$ the values of $\sigma(\alpha)$ are exactly the vector coordinates $\in \mathbf{R}^2$ corresponding to the $\alpha$-cuts $C_{\alpha}\bigl(tr^*(2,4,5)\bigr).$
\end{remark}

This leads to the following way of looking at things, a representation of a fuzzy triangle number as a projected parameterized curve $\sigma(\alpha)$ in $\R^2$ (obviously the curvature $\kappa$ of $\sigma$ is constantly $\kappa = 0$, and the projected curve appears as a straight line):

\begin{figure}[H]
  \centering
  \includegraphics[width=7cm]{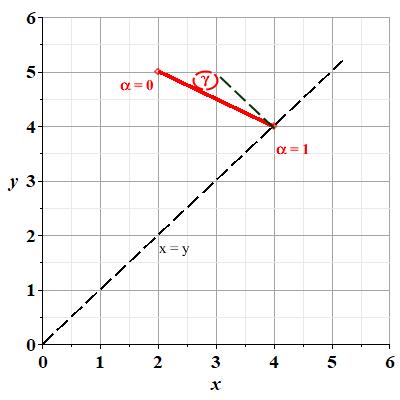}
  \caption{The middle point $x = 4$ of the triangle number $tr^*(2,4,5)$ corresponds to $(4,4)$ lying on the $\{x=y\}$ - line (with $\alpha = 1$ implicit) whereas the sides $l=2$ and $r=5$ appear as the other endpoint $(2,5)$ of the parameterized curve with $\alpha = 0$ implicit.}\label{FigureExample1by3dProjection}
\end{figure}

Because $\sigma(\alpha)$ is linear in all coordinates it has a single unique tangent vector $T_{\xi} = \sigma'(\alpha)$ given by
\begin{equation}
\sigma'(\alpha) = \langle m-l,\,-r+m \rangle = \langle 2,\,-1 \rangle = - \langle -2,\,1 \rangle
\end{equation} which together with the boundary condition $\xi(m) = 1,$ that is knowledge of the location of the middle point $m,$ uniquely determines the fuzzy triangle, as shown in \textit{Fig.~\ref{FigureExample1TangentAngle}}:
\begin{figure}[H]
  \centering
  \includegraphics[width=7cm]{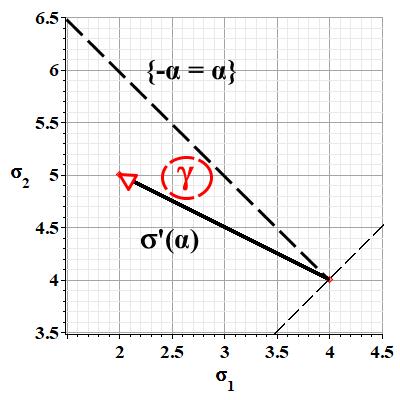}
  \caption{The free tangent vector $\sigma'(\alpha) = \langle \, - (m-l),\, r-m \,\rangle$ attached to the triangle's middle point $(m,m)$ (for better visualization we take the vector's negative against the run of the parametrisation).}\label{FigureExample1TangentAngle}
\end{figure}

Then we note that
\begin{itemize}

\item
$T_{\xi}$ = $\sigma'(\alpha)$ is of constant magnitude
\begin{equation}\label{EquationExampleTriangleMagnitude}
|\sigma'(\alpha)| = \sqrt{\left((m-l)^2+(r-m)^2\right)},
\end{equation}\sp

which in the example case $\xi = tr^*(2,4,5)$ gives
$$ = \sqrt{(-2)^2+1^2} = \sqrt{5}.$$

\item
The constant angle $\gamma$ by which the curve's tangent vector deviates from the line $\{-\alpha = \alpha\}$ perpendicular to the $\{x = y\}$ - line is
\begin{equation}\label{EquationGammaAngleTriangleDerivation}
\gamma_{\xi}(\alpha) = \pm\,\arccos\frac{\langle\sigma'(\alpha),(1,-1)\rangle } {|\sigma'(\alpha)|\cdot|(1,-1|} =
\end{equation}
which simplifies to
\begin{equation} \label{EquationGammaArcCosine}
= \pm\,\arccos\left(\frac{1}{\sqrt{2}}\cdot\frac{(r-l)}{|\mathbf{r}|}\right)
\end{equation}
with
\begin{equation}
= 0 \quad \text{ for} \quad r-m = l-m,
\end{equation}
and
\begin{equation}
\begin{cases}
+  \text{ if }\quad m-l < r-m, \\
-  \text{ if }\quad m-l > r-m,
\end{cases}
\end{equation}
as anticipated in~\eqref{EquationGammaTriangle}.
\end{itemize}

\noindent Note that the for triangle numbers of the sort,

\begin{align}
\xi &= tr^*(l,r,r) \\
& \text{ or } \\
\xi &= tr^*(l,l,r)
\end{align}

\noindent that is of extreme skew, \eqref{EquationGammaArcCosine} produces an angle of

\begin{equation}
\gamma =
\begin{cases}
\qquad - 45^{\circ} \\
\qquad \,\, \text{ or } \\
\qquad + 45^{\circ},
\end{cases}
\end{equation}
respectively. \sp

In this particular example $\xi = tr^*(2,4,5)$ the constant angle of deviation $\gamma(\alpha)$ is seen to be given by
\begin{equation}
 cos (\gamma^c) = 2/\sqrt{5}
\end{equation}
where $\gamma^c + \gamma = 45^\circ$ that is to say $\gamma = \arccos(2/\sqrt{5})  - 45^\circ = - 18.435$ degrees. \sp

\bigskip
\noindent For the full picture the length of the projected curve is by definition
\begin{equation}\label{EquationExampleTriangleLength}
\mathbf{C}(\sigma) = \int_{0}^{1}|\sigma'(\alpha)|\,d\alpha = \sqrt{2^2 + 1^2} = \sqrt{5},
\end{equation}
\noindent which can, in this linear case, be seen by the Pythagoras theorem.

\begin{definition}
It is this angle $\mathbf{\gamma}_{\xi}$ \eqref{EquationGammaAngleTriangleDerivation} by which the curve's tangent vector deviates from the $\{\alpha = \alpha\}$-axis, that we take as skewness measure for fuzzy triangle numbers.
\end{definition}

To see why this choice makes sense as a measure of deviation from symmetry it is enough to realize that the projected parameterized curve as well the defining tangent vector representing a symmetric triangle number lies entirely on the $\{-\alpha = \alpha\}$ line, that is $\gamma = 0$ for symmetric fuzzy triangle numbers (postulate P.2). This is visualized below for three fuzzy triangle numbers: $tr^*(2,4,5)$, its reflection $tr^*(3,4,6)$ and the symmetric triangle $tr^*(2.5,4,5.5)$:

\subsubsection{Skewness as angle of deviation from symmetry}
\begin{figure}[H]
\begin{subfigure}{.33\textwidth}
  \centering
  \includegraphics[width=.8\linewidth]{Example1TangentAngle}
  \caption{left skew}
  \label{fig:sfig1}
\end{subfigure}%
\begin{subfigure}{.33\textwidth}
  \centering
  \includegraphics[width=.8\linewidth]{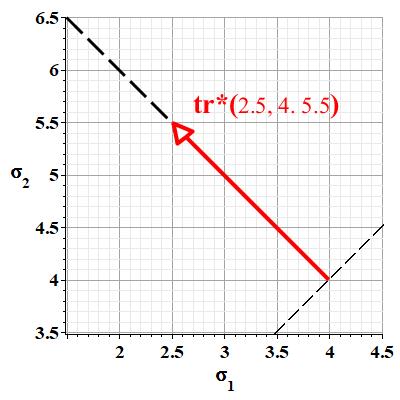}
  \caption{symmetry}
  \label{fig:sfig2}
\end{subfigure}
\begin{subfigure}{.33\textwidth}
  \centering
  \includegraphics[width=.8\linewidth]{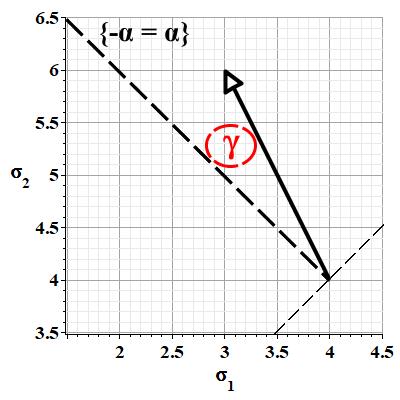}
  \caption{right skew}
  \label{fig:sfig3}
\end{subfigure}
 \caption{The tangent vector $\vec{(-2,1)}$ representing the fuzzy triangle number $tr^*(2,4,5)$ \ref{fig:sfig1} deviates from symmetry by $\gamma \sim - 18,45^\circ$. The tangent vector $\vec{(-1.5,1.5)}$ representing the symmetric triangle $tr^*(2.5,\,4,\,5.5)$ \ref{fig:sfig2} has deviation $\gamma = 0$ and the reflected triangle number $tr^*(3,4,6)$ \ref{fig:sfig3} represented by the tangent vector $\vec{(-1,2)}$ deviates from symmetry by an angle of $\gamma \sim + 18.45^\circ$.}\label{FigureTangentAngleDeviations}
\end{figure}

The next observation is that the vector $\sigma'(\alpha),$ as any vector, may be given in polar rather than Cartesian coordinates, by setting
\begin{align}\label{EquationTrianglePolar}
\mathbf{r}_{\xi} & = |T_{\xi}|, \\
\gamma_{\xi} &= \text{ as in \eqref{EquationGammaTriangle}}.
\end{align}

The next logical step is to use this \eqref{EquationTrianglePolar} as an alternative (eventually primary) representation of any fuzzy triangle number, whereby it is visually more convenient to take $\gamma$ from verticality :

\subsection{Polar coordinates. Tangent vector and middle point as a viable representation of a fuzzy triangle number}$ $\sp

\noindent Following the 1-1 correspondence between the ordered triples of \sp

\begin{equation}\label{EquationCorrespondencePolarAndTriangle}
(m,\mathbf{r},\gamma) \Leftrightarrow (l,m,r)
\end{equation}
for
\begin{equation}
l \leq m \leq r \in\R \quad \text{ and } \quad \gamma\in\left[-\frac{\pi}{4},\frac{\pi}{4}\right],\, \mathbf{r}\in[0,\infty)
\end{equation}
one may rewrite a linear fuzzy number $\xi$ in \emph{polar representation} as

\begin{equation}
\xi = \left(m,\mathbf{r},\gamma\right)
\end{equation}

\begin{figure}[H]
\begin{subfigure}{.40\textwidth}
  \centering
  \includegraphics[width=0.9\linewidth]{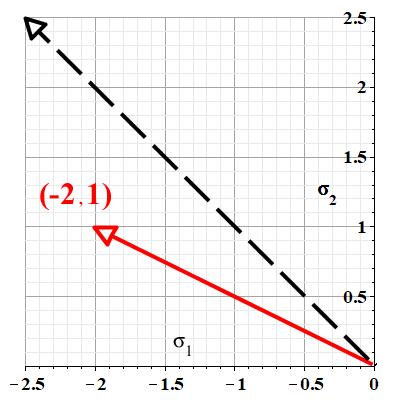}\caption{Cartesian coordinates}
\end{subfigure}
\begin{subfigure}{.50\textwidth}
\centering
  \includegraphics[width=0.9\linewidth]{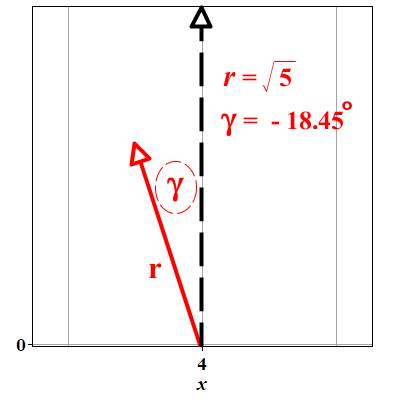}\caption{Polar coordinates}
\end{subfigure}
  \caption{The tangent vector representing the fuzzy triangle number $tr^*(2,4,5)$ in Cartesian and polar coordinates}\label{FigureExample1TangentAngle0}
\end{figure}

\subsection{A change of basis for better visualization}

The visual change from cartesian to polar coordinates counting $\gamma$ of the ordinate axis and the characterizing vector's tail anchored on the abscissa, in the real number $m$ which is being fuzzified, may be effected by a linear transformation, which amounts to a change of basis namely by
\begin{equation}\label{EquationMatrixF}
\mathbf{F} = \frac{1}{2}\cdot
\left(
  \begin{array}{cc}
    1 & 1 \\
    -1 & 1 \\
  \end{array} ,
\right)
\end{equation}
which takes
$$\binom{1}{1} \rightarrow \binom{1}{0}$$
and
$$\binom{-1}{1}\rightarrow\binom{0}{1}$$
making the $\{x=y\}$ axis the new abscissa and $\{-x=y\}$ the new ordinate axis scaling both \mbox{by $\dfrac{1}{\sqrt{2}}.$} \sp

Note that $\mathbf{F}$ is an orthogonal transformation which leaves the angles $\gamma$ unchanged. We use this transformation in the next section to better visualize our line of reasoning for the one-to-one correspondence of non-linear fuzzy membership functions and a certain class of parameterized curves. \sp

We finish this section with a comparative illustration of fuzzy triangle numbers in traditional versus vector representation, which also serves to show van Zwet's $P.1 - P.3.$ The formal, easy proofs of those are shown in generality, that is for linear and non-linear fuzzy numbers in the next section,~\ref{PropertiesVanZwet}.

Note that all illustrations below are given in the $\mathbf{F}$-coordinate system, that is vector lengths $|\mathbf{r}|$ are scaled by $\dfrac{1}{\sqrt{2}}.$

\begin{figure}[H]
  \centering
   \includegraphics[width=3.5cm]{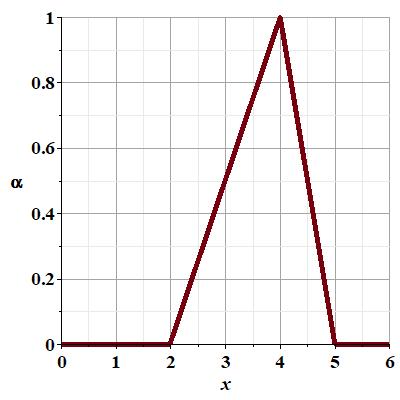}
   \includegraphics[width=3.5cm]{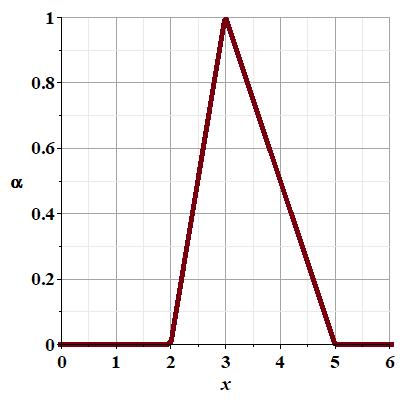}
   \includegraphics[width=3.5cm]{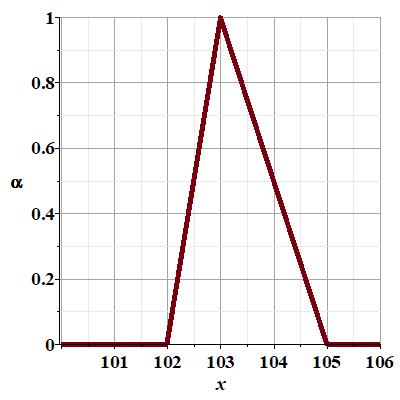} \\
   $ $
   \includegraphics[width=3.5cm]{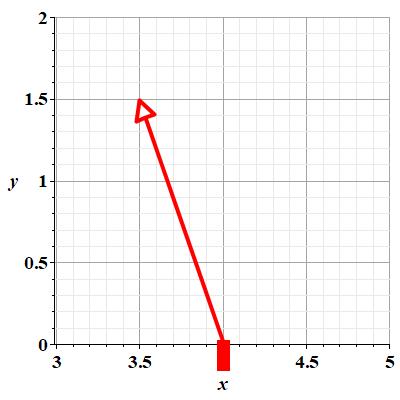}
   \includegraphics[width=3.5cm]{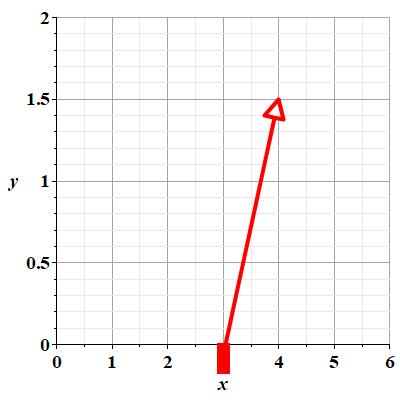}
   \includegraphics[width=3.5cm]{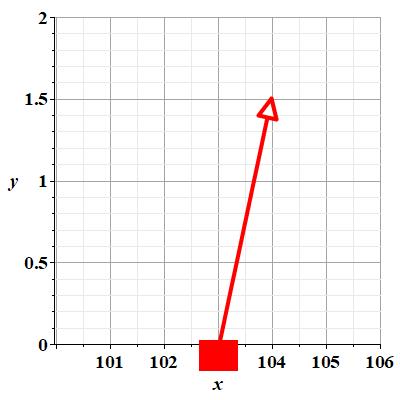}
\caption{Example fuzzy triangle numbers and their vector counterparts. Antisymmetry and translation invariance.}\label{FigureExampleTriangleVector}
\end{figure}

\begin{figure}[H]
  \centering
 \includegraphics[width=3.5cm]{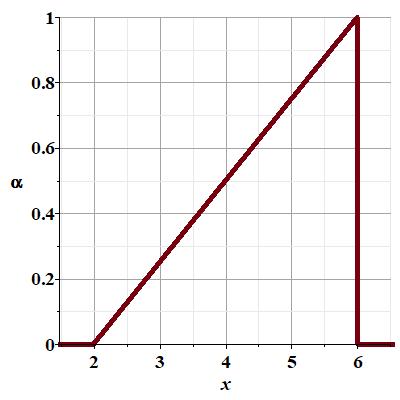}
 \includegraphics[width=3.5cm]{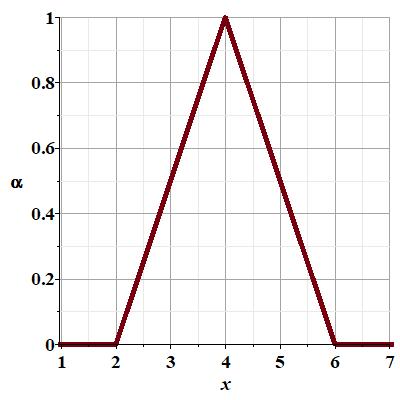}
 \includegraphics[width=3.5cm]{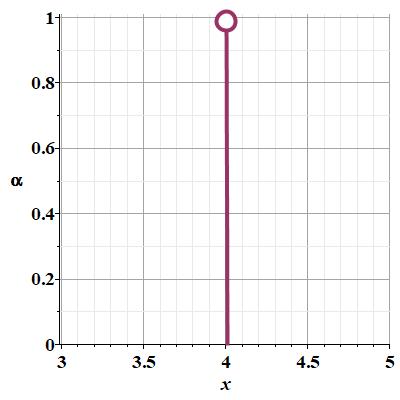} \\
 $ $
 \includegraphics[width=3.5cm]{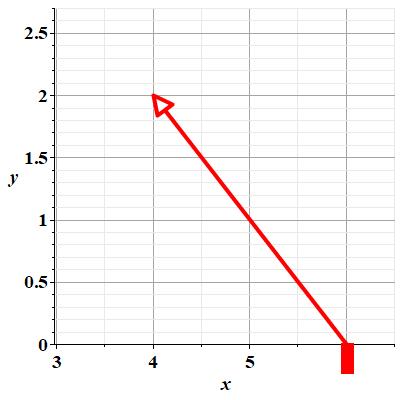}
 \includegraphics[width=3.5cm]{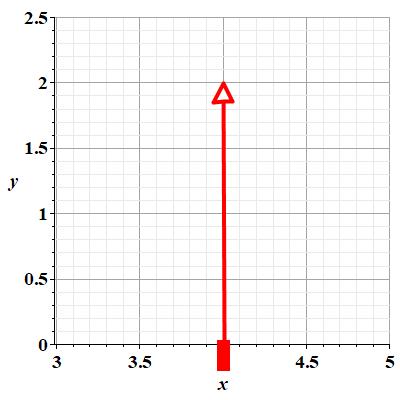}
 \includegraphics[width=3.5cm]{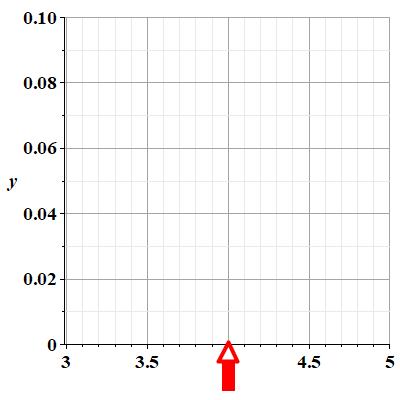}
 \caption{Special (border) cases: Triangle numbers of extreme skewness i.e. $l=m$, $m=r$, Symmetric triangle numbers ($m-l = r-m$).  and exact, real numbers i.e. $l=m=r$.}\label{FigureExampleTriangleVectorBorderCases}
\end{figure}

\noindent Note that
\begin{enumerate}
\item symmetric triangle numbers $tr^*(m-\delta,m,m+\delta)$ translate into vertical arrows of length (height)
$\mathbf{r} = \sqrt{2}\,\delta,$ and angle $\gamma=0,$
\item real (exact) numbers translate into themselves on the real line ($r = 0, \gamma = 0$),
\item and triangle numbers of type $tr^*(l,m,m)$ or $tr^*(m,r,r)$ translate into vectors of length $(m-l)$, respectively $(r-m)$ and angle $\gamma = \pm 45^\circ.$
\end{enumerate}

\section{Non-linear fuzzy numbers in polar coordinates}

For the purpose of this section we assume for a fuzzy number $\xi = \left(l(x),r(x)\right)$ to hold some additional differentiability properties:

\begin{properties}[of a non-linear fuzzy number]\nonumber
As had: $l(x),r(x)$ are supported on $[l,m], [m,r]$ respectively, and onto $[0,1],$ invertible and such that $d(\alpha) = l^{-1}(\alpha),$
and now we also assume: $d(\alpha),\,u(\alpha)$ to be continuously differentiable on $[0,1].$ \sp

We also make the following \textbf{assumptions} whose sense becomes immediate below
\begin{equation}\label{assupmtionFN1}
0 < d'(\alpha) < \infty, \text{ ($d$ strictly increasing and $d'$ bounded)}
\end{equation}
\begin{equation}\label{assupmtionFN2}
0 < u'(\alpha) < \infty. \text{ ($u$ strictly increasing and $u'$ bounded)}
\end{equation}
\end{properties}

\begin{remark}\label{RemarkSobolev}
Much less restrictive assumptions may be adopted and still achieve all results of this paper, such as: non-strict monotonicity of $l$ and $r$ instead of strict, semi-continuity instead of continuous differentiability, support unbounded but with \newline $\lim_{x \rightarrow \infty} l(x) = 0$ and $\lim_{x \rightarrow \infty} r(x) = 0$ instead of bounded support. To overcome the appearing technical obstacles one would venture into the theory of distributions (in the sense of Laurent Schwartz, not probability distributions) but maximum generalization is not a concern at this point.
\end{remark}

We now repeat step by step for non-linear numbers the line of reasoning taken in section~\ref{SectionTriangles} for fuzzy triangle (linear) numbers:

\subsection{Derivation}\label{SubsectionDerivationNonLinear}$ $\par

\noindent Given a fuzzy number as
\begin{multline}\label{RepresentationX}
\left(l(x),r(x)\right), {x\in\R} \hfill \text{\footnotesize(TRADITIONAL REPRESENTATION)}
\end{multline}
we may first switch to
\begin{multline}\label{RepresentationAlpha}
\left[d(\alpha),u(\alpha)\right], \alpha\in[0,1] \hfill \text{\footnotesize(PARAMETRIC REPRESENTATION)}
\end{multline}
which one may choose to understand as a parameterized curve
\begin{multline}\label{RepresentationCurve}
\sigma(\alpha):=\,\left(d(\alpha),u(\alpha)\right), \quad \alpha\in[0,1]. \qquad \text{\footnotesize(AS A PARAMETERIZED CURVE)}
\end{multline}

\noindent By assumptions \eqref{assupmtionFN1}, \eqref{assupmtionFN2} one may equivalently give \eqref{RepresentationCurve} as
\begin{multline}\label{RepresentationTangent}
\xi = \sigma'(\alpha) = \left(d'(\alpha),u'(\alpha)\right) \text{ with boundary values } \left(d(1),u(1)\right) = (m,m),\\
 \text{\footnotesize(AS TANGENT BUNDLE)}.
\end{multline}

\noindent For convenient visualization one may choose to instate a change of coordinates by $\mathbf{F}$ (see~\eqref{EquationMatrixF}) as done below in figure~\ref{FigureFourLeafClover}.

\begin{figure}[H]
\begin{subfigure}{.33\textwidth}
  \centering
  \includegraphics[width=.8\linewidth]{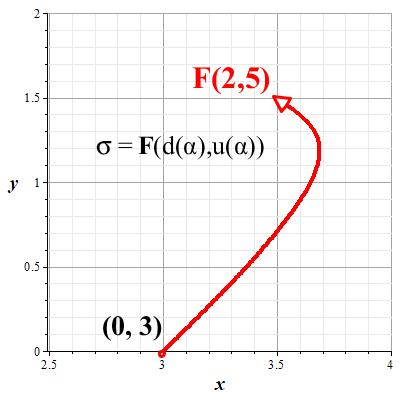}
  \caption{The curve $\mathbf{F}(\sigma)(\alpha)$}
  \label{fig:(s2fig1)}
\end{subfigure}%
\begin{subfigure}{.33\textwidth}
  \centering
  \includegraphics[width=.8\linewidth]{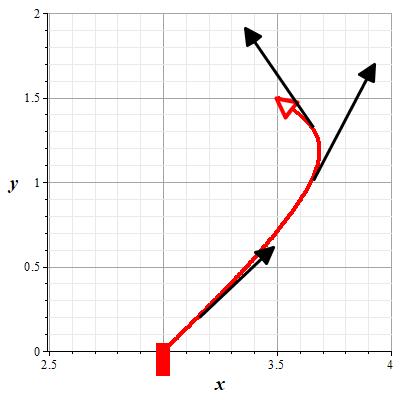}
  \caption{$ $}
  \label{fig:s2fig2}
\end{subfigure}
\begin{subfigure}{.33\textwidth}
  \centering
  \includegraphics[width=.8\linewidth]{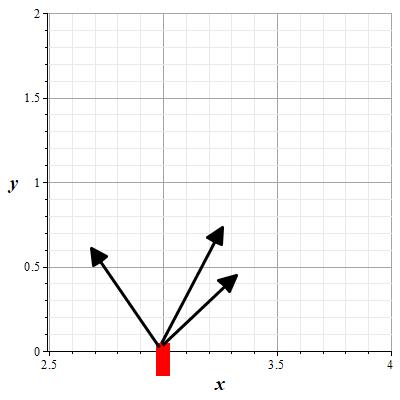}
  \caption{The tangent bundle}
  \label{fig:s2fig3}
\end{subfigure}
 \caption{A non-linear fuzzy number supported on $\mathbf{F^{-1}}\binom{3.5}{1.5} = [2,5]$ with middle point $m=3$ as an $\mathbf{F}$-transformed parameterized curve~\ref{fig:(s2fig1)}, the same curve with indicated tangent vectors, and as a tangent bundle.}\label{FigureFourLeafClover}
\end{figure}
$\diamond$\sp
The \textbf{key} to all is again to represent $\xi$ in polar coordinates:\sp

\noindent As done before in the linear case one may change coordinates to polar and rewrite each tangent vector for each $\alpha$ as the triple

\begin{equation}\label{RepresentationPseudoPolar}
\xi = \left(m,\mathbf{r}(\alpha),\gamma(\alpha)\right)
\end{equation}

where as before

\begin{equation}
\mathbf{r}(\alpha) = |\sigma'(\alpha)| = \sqrt{d'(\alpha)^2 + u'(\alpha)^2}.
\end{equation}\sp

and
\begin{multline}\label{EquationGammaGeneral}
\gamma(\alpha):= \pm\arccos\left(\frac{<(d'(\alpha),u'(\alpha)),(1,-1)>}{\mid\sqrt{d'(\alpha)^2 + u'(\alpha)^2}\mid\cdot\mid(1,-1)\mid}\right) = \\
= \pm\arccos\left(\frac{u'(\alpha) - d'(\alpha)}{\sqrt{2}\:\mathbf{r}}\right).
\end{multline}
with the sign determined as in \eqref{EquationGammaTriangleSign} by

\begin{equation}\label{EquationGammaGeneralSign}
\begin{cases}
+ \quad for \quad \abs{d'}<\abs{u'} \\
- \quad  for \quad \abs{d'}>\abs{u'} \\
0 \quad  for \quad \abs{d'}=\abs{u'} \qquad \text{(symmetry)}
\end{cases}
\end{equation}

with sign changes taking place at $\alpha$ for which $\gamma(\alpha) = 0.$

\begin{remark}
Differentiation~\eqref{EquationGammaGeneralSign} is redundant and artificial when the principle values of cosine to be taken as $[-\pi,\pi]$ instead of $[0,\pi]$ as is custom for the majority of textbooks and calculators / computer programs.
\end{remark}

The re-transformation from polar to cartesian runs:

\begin{equation}\label{EquationBacktransformationTriangle}
\begin{cases}
\mathbf{r}(\sin(\gamma)\,\pm\frac{\pi}{4}) &\rightarrow \, m-l, \\
\mathbf{r}(\cos(\gamma)\,\pm\frac{\pi}{4}) &\rightarrow \, r-m,
\end{cases}
\end{equation}

With the sign determined by reversing~\eqref{EquationGammaTriangleSign}. This transformation is clear from \textit{Fig.~\eqref{FigureExample1TangentAngle0}}.

\Links
\subsection{A numerical example}\noindent\begin{example}\label{Example3NonLinear}
Let $\xi$ be the fuzzy number around $m=\pi$ whose sides are given in traditional representation by the ordered function pair $\xi = \left(l(x),r(x)\right):$ \sp

\noindent
\begin{equation}
\begin{cases}
l(x) &= -1/3+\arccos\left(\sqrt{(\cos(4/3)^2+\pi-x)}\right) \text { on }  [2.30398574,\pi] \\
r(x) &= (2-(1/\pi)\cdot x)^{1/4} \qquad \text { on }  [\pi,2\pi]
\end{cases}
\end{equation}

\begin{figure}[H]
  \centering
  \includegraphics[width=12cm]{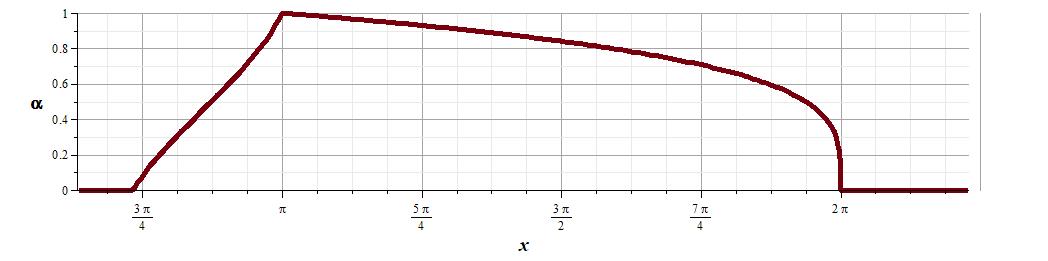}
  \caption{The membership function of example \eqref{Example3NonLinear} in traditional representation}\label{FigureExample3oneFunction}
\end{figure}

\noindent equivalently in parametric representation $\alpha\in[0,1]$
\noindent

$$\xi = [d(\alpha),u(\alpha)],$$ with
\begin{equation}
\begin{cases}
d(\alpha) &= \pi+\left(\cos(1+1/3)\right)^2-\left(\cos(\alpha+1/3)\right)^2, \\
u(\alpha) &= -\pi\alpha^4+2\cdot\pi,
\end{cases}
\end{equation}
$ $\sp

\noindent as tangent vector bundle
$\sigma'(\alpha) = \left[
  \begin{array}{c}
    d'(\alpha) \\
    u'(\alpha) \\
  \end{array}
\right],$
where
\begin{equation}
\begin{cases}
d'(\alpha) &= 2\cos(\alpha+1/3)\sin(\alpha+1/3), \\
u'(\alpha) &= 4\pi\alpha^3
\end{cases}
\end{equation}

\Rechts
\noindent and finally in polar coordinates as $\left(\mathbf{r}(\alpha),\gamma(\alpha)\right)$ with
\begin{align}
\mathbf{r}(\alpha) &= 2\,\sqrt{\cos(\alpha+1/3)^2\cdot\sin(\alpha+1/3)^2+4\pi^2\alpha^6} \end{align} and
\Rechts
\begin{equation}\label{EquationExample3Gamma}
\gamma(\alpha) =
\pm 1/2\,{\frac {\sqrt {2} \left( 2\,\pi \,{\alpha}^{3}+\cos \left( \alpha+1/3
 \right) \sin \left( \alpha+1/3 \right)  \right) }{\sqrt {4\,{\pi }^{2}{\alpha}^
{6}- \left( \cos \left( \alpha+1/3 \right)  \right) ^{4}+ \left( \cos
 \left( \alpha+1/3 \right)  \right) ^{2}}}}
\end{equation} \sp

\noindent with a (single) sign change taking place at $\alpha = 0.4299872156.$
\end{example}

The parameterized curve and its polar coordinate functions are displayed below in \textit{Fig.~\ref{FigureExample3gammaAndradius}}:
\begin{figure}[H]
  \centering
  \includegraphics[width=5.5cm]{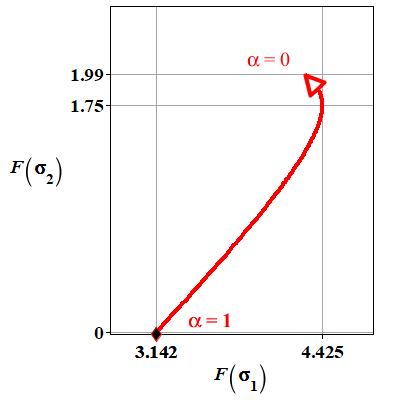}
  \includegraphics[height=6cm]{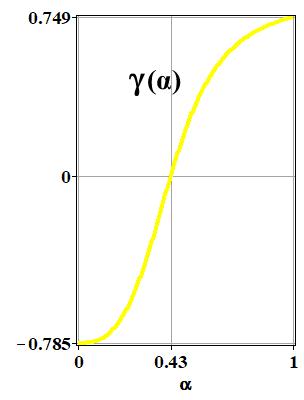}
  \includegraphics[height=6cm]{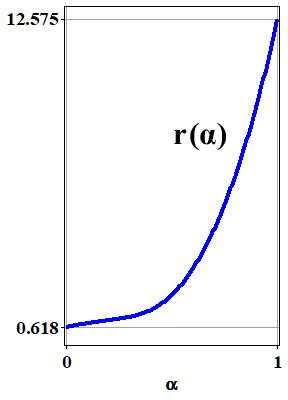}
  \caption{The fuzzy number $\xi$ represented as an $\mathbf{F}$-transformed parameterized curve, and its pseudo-polar coordinate functions $r(\alpha)$ and $\gamma(\alpha).$ Note how $\gamma(\alpha)$ is contained between $\pm\,\frac{\pi}{4} \sim 0.7854.$ Also mind the direction of flow from $\alpha = 0$ to $\alpha = 1$.}\label{FigureExample3gammaAndradius}
\end{figure}

We are now equipped to give our definitions of skewness, mean value and dispersion of non-linear fuzzy numbers.

\section{Skewness of non-linear fuzzy numbers}\label{SectionSkewness}

\noindent When looking at~\eqref{RepresentationPseudoPolar} the following definition of skewness very strongly suggests itself: \sp

\subsection{Definition}
\begin{definition}[Skewness at a point, $\alpha$-level]$ $\sp

\begin{equation}\label{EquationSkewnessFuzzyNumberPoint}
S^*(\xi)_{\alpha_0} = \gamma(\alpha_0) = \pm \arccos\left(\frac{d'(\alpha_0) - u'(\alpha_0)}{\sqrt{2}\,\mathbf{r}}\right).
\end{equation}

\end{definition}

By design $-\frac{\pi}{4}\leq S^*(\xi)_{\alpha_0} \leq \frac{\pi}{4},$ whereby both extreme values may be attained multiply in a single fuzzy number, though in most typical cases there will be one single sign-change, or none at all.\sp

It is very valuable for practical applications to have skewness at a given $\alpha$-level defined. The overall skewness coefficient as defined below in~\eqref{EquationSkewnessFuzzyNumberOverall} is eigen to infinitely many very different membership functions, i. e. does not reflect the characteristics of a concrete FN which stands for the very concrete value of a concrete linguistic variable.\sp
For practical purposes any given FN may be approximated with deliberate accuracy by the appropriate (finite) amount of $\alpha$-cuts. (This can be shown in different ways but is most easily seen when regarding a fuzzy number as a Lebesgue integral achieved by simple functions defined by these $\alpha$-cuts).
The huge gain from having skewness coefficients defined on all $\alpha$-levels, instead of having only one, overall measure, is apparent when an optimization problem $\mathbf{P^*}$ involving skewness is formulated in fuzzy variables and/or parameters.
The problem becomes tractable by choosing a partition of $[0,1]$, levels of special interest, according to problem specific priorities or evenly spaced, and then to receive a number of problems in real variables and/or parameters. \sp
If that is, say $n$, cuts, then the computational complexity of a linear program increases by the multiplicative factor $n$, with the characteristics of the FN reflected to whatever level chosen.\sp
Having a skewness coefficient at every point $\alpha_0$ along the curve also allows to treat each $\alpha_0$-level $[d(\alpha_0),u(\alpha_0)]$ as an ordered pair of interval and skewness coefficient

\begin{equation}\label{EquationAlphaCutWithSkewness}
[d(\alpha_0),u(\alpha_0)],\,\gamma(\alpha_0).
\end{equation}\sp

Anyhow the overall skewness coefficient of a given fuzzy number \underline{must} then be defined as

\begin{definition}[Overall skewness of a fuzzy number]\label{DefinitionSkewnessFuzzyNumber}$ $\sp

\begin{equation}\label{EquationSkewnessFuzzyNumberOverall}
S ^*(\xi) = \dfrac{1}{C(\sigma)} \int_0^1 \gamma(\alpha)\cdot\mid\sigma'(\alpha)\mid\,d\,\alpha ,
\end{equation}
\end{definition}

\noindent where

\begin{equation}\label{EquationLenghtOfCurve}
C(\sigma):=\int_0^1 \mid\sigma'(\alpha)\mid\,d\,\alpha
\end{equation}

\noindent is the length of the parameterized curve $\sigma.$\sp

\bigskip
The values of both skewness coefficients, $S ^*(\xi)$ and $S ^*(\xi)_{\alpha_0}$ are contained in the interval
$$\gamma \in \left[-\frac{\pi}{4},\,\frac{\pi}{4}\right].$$ \sp

\begin{remark}
For some a range of values of $\left[-\frac{\pi}{2},\,\frac{\pi}{2}\right]$ may seem more desirable for intuition or computability. This may of course be achieved without distortion and loss of information by scaling.
\end{remark}

\subsection{Properties (van Zwet 1964)}\label{PropertiesVanZwet}
It is straightforward to see that van Zwet's three (of four) conditions for a ``useful'' skewness measure are fulfilled:

\begin{enumerate}
\item[\textbf{P.1}] A scale or location change for a fuzzy number does not alter $\gamma.$ \sp

In the context of fuzzy numbers this reads:
\begin{equation}
\text{ if } \nu=c\,\xi+d \text{ then } \gamma(\nu)=\gamma(\xi).
\end{equation}

This is clear since:\sp

The location parameter $d$ is annulled by differentiation
\Links
\begin{equation}
 \left(c\cdot\sigma(\alpha)+d\right)' = c\cdot\sigma'(\alpha)
\end{equation}
and the scaling parameter also does not change the angle as
\begin{equation}
 \frac{<(c d'(\alpha),c u'(\alpha)),(1,-1)>}{\mid\sqrt{c^2 d'(\alpha)^2 + c^2 u'(\alpha)^2}\mid\cdot\mid(1,-1)\mid} = \frac{<(d'(\alpha),u'(\alpha)),(1,-1)>}{\mid\sqrt{d'(\alpha)^2 + u'(\alpha)^2}\mid\cdot\mid(1,-1)\mid}
\end{equation}

\item[\textbf{P.2}] For a symmetric distribution $\gamma = 0.$ \sp

Symmetricity of a fuzzy number and when writing in parametric representation implies
\Links
\begin{equation}
u'(\alpha) = - d'(\alpha)
\end{equation}
so the property follows immediately. See \textit{Fig.~\ref{FigureTangentAngleDeviations}}.\sp

\item[\textbf{P.3}] If $Y=-X$ then $\gamma(Y)=-\gamma(X).$ \sp

In the context of fuzzy numbers this is to be interpreted as a reflection around some axis $x = s$,
Set in parametric representation $X = [d,\,u]$ then $Y= [-u,\,-d] + s$ and the property
follows directly from~\eqref{EquationGammaGeneralSign}. \sp

\end{enumerate}$ $\sp

\noindent Returning to \textbf{example~\ref{Example3NonLinear}}, that is the fuzzy number $\xi$ shown in \textit{Fig.~\ref{FigureExample3oneFunction}} given in parametric representation by

\begin{equation}
\xi =
\begin{cases}
d(\alpha) &= \pi+\left(\cos(1+1/3)\right)^2-\left(\cos(\alpha+1/3)\right)^2, \\
u(\alpha) &= -\pi\alpha^4+2\cdot\pi,
\end{cases}
\end{equation}
$ $\sp

\noindent its point skewness is given by $\gamma(\alpha)$ of its polar representation~\eqref{EquationExample3Gamma}, i.e.:

\begin{equation}
S^*(\xi)_{\alpha} = \pm 1/2\,{\frac {\sqrt {2} \left( 2\,\pi \,{\alpha}^{3}+\cos \left( \alpha+1/3
 \right) \sin \left( \alpha+1/3 \right)  \right) }{\sqrt {4\,{\pi }^{2}{\alpha}^
{6}- \left( \cos \left( \alpha+1/3 \right)  \right) ^{4}+ \left( \cos
 \left( \alpha+1/3 \right)  \right) ^{2}}}}
\end{equation} \sp

\noindent with a (single) sign change taking place at $\alpha = 0.4299872156.$ \sp

\noindent To compute the overall skewness coefficient of $\xi$ by \eqref{EquationSkewnessFuzzyNumberOverall} we need compute the length $\mathbf{C}(\sigma)$ of the curve $\sigma(\alpha)$ by \eqref{EquationLenghtOfCurve}:

\begin{equation}
\mathbf{C}(\sigma) = \int_{0}^{1}|\sigma'(\alpha)|\,d\alpha = \int_{0}^{1}\mathbf{r}(\alpha)\,d\alpha = 3.503009852.
\end{equation}\sp

\noindent The overall skewness coefficient of $\xi$ is thus
\begin{multline}
S^*(\xi) = \dfrac{1}{C(\sigma)} \int_0^1 \gamma(\alpha)\cdot\mid\sigma'(\alpha)\mid\,d\,\alpha  = \frac{1}{3.503009852}\cdot 1.760449519 \\$ $\\
= 0.5025534021\,[rad] = 0.1599677162\,\pi.
\end{multline}

\section{Mean by mean value theorem for integrals}\label{SectionMean}
In this section we offer a contribution to the already rich literature on what may be termed a mean value of a fuzzy number.\sp

Although fuzzy arithmetic is very straightforward (once given in parametric representation) and under certain assumptions also functional calculus is theoretically well founded the main drawback remains computational complexity, and in a systematic sense that the lack of an inverse element~\cite{BKKN97} (multiplicative or additive) typically leaves functional equations involving fuzzy numbers without a solution if understood as equations in the usual ``$\,=\,$'' way and not some other, weaker relation. \sp

Hence the quest for a set of descriptive parameters which, to a satisfactory degree, reflect the characteristic basic attributes contained in a given fuzzy number while alleviating aforementioned difficulties.\sp

The first thing that comes to mind is naturally that of a ``mean'' or ``expected'' value, and the earliest days the precursors of fuzzy set theory have set out to define appropriate measures:\sp
One may categorize three different approaches (strategies): \sp

\textbf{1.} a single (one-dimensional) mean value.
As an example: the possibilistic mean value of Goetschel and Voxman~\cite{GV86} has proven to be particularly useful over time, as is its generalization by Carlson and Full\'{e}r~\cite{CF01}. This mean value also appears appears naturally when setting a density $\rho(x) = \mu(x)\mu'(x)$ in an effort to relate directly to the machinery of probability as in \cite{LGY15}, (see \ref{SubsubsectionMomentProbabilityFuzzy}).\sp
Really every single so called ranking function (index, method) may play the role of a mean value. The reader is referred to~\cite{BM13} for a comprehensive comparative treatment of the wide variety of ranking indices and methods.\sp

\textbf{2.} An interval valued mean value\sp
which may show a relation to probability theory as by Didier Dubois and Henri Prade~\cite{DP87} (who refer to Dempster~\cite{Dem67}), or without such relation as ``On possibilistic mean value and variance of fuzzy numbers'' by  R. Full\'{e}r, P. Majlender~\cite{FM03}, who
define an lower and upper possibilistic mean building on~\cite{GV86}. \sp

\textbf{3.} A  third approach is to find a canonical most representative, triangle or trapezoidal fuzzy number (i.e. sets of three resp. four real numbers which are closed with respect to addition) which shares the same characteristic parameters (such as any of the ranking indices listed in~\cite{BM13} or later, or ``value'' and ``ambiguity'' introduced in~\cite{DVW98}, or fuzziness introduced in~\cite{DVW298}), or is closest to it with respect to a chosen metric (see e.g~\cite{DK99} as a staring point) or other criterion. \par

Having already stated \eqref{EquationAlphaCutWithSkewness} as a variation of approach \textbf{2.} above, in this paper we take the third approach:\sp

Having defined the overall skewness coefficient $\gamma$ to a given FN we may define a ``\emph{mean value triangle}'' i.e. a linear (triangle)  number ``most representative'' of the non-linear FN in the sense that it shares, as the characteristic parameter, the same overall skewness coefficient and the same middle point $m$.\sp

We start with an interesting representation theorem which needs no separate proof, as it follows directly from the preceding deliberations and definitions:

\begin{theorem}
Following the 1-1 correspondence between ordered triples of
\begin{equation}\label{EquationCorrespondencePolarAndTriangleFunction}
(m,\mathbf{r}(\alpha),\gamma(\alpha)) \Leftrightarrow tr^*(l(\alpha),m,r(\alpha))
\end{equation}
for
\begin{equation}
l(\alpha) \leq m \leq r(\alpha) \in\R \quad \text{ and } \quad \gamma\in\left[-\frac{\pi}{4},\frac{\pi}{4}\right],\, \mathbf{r}\in[0,\infty)
\end{equation}

any given fuzzy number satisfying at least~\eqref{assupmtionFN1},~\eqref{assupmtionFN2}  may be represented by an ordered set of triangle fuzzy numbers given by~\eqref{EquationCorrespondencePolarAndTriangleFunction}.
\end{theorem}

Now we make use of the common mean value theorem for integrals, that is we find the value $\alpha_{mean} \in [0,1]$ at which level the overall skewness coefficient of the measured fuzzy number is attained:

\begin{equation}\label{EquationMeanAlphaValue}
\exists\, \alpha_{mean}: \gamma(\alpha_{mean}) = \dfrac{1}{C(\sigma)} \int_0^1 \gamma(\alpha)\cdot\mid\sigma'(\alpha)\mid\,d\,\alpha.
\end{equation}

Having found $\alpha_{mean}$ gives us the pseudo-polar coordinates of the corresponding tangent vector to the curve at $\alpha_{mean}$:

\begin{equation}\label{EquationMeanValueTrianglePolar}
\left(\mathbf{r}(\alpha_{mean}),\gamma(\alpha_{mean})\right),
\end{equation}
and in consequence by \eqref{EquationBacktransformationTriangle} a fuzzy triangle number $tr^*(l,m,r)$ in traditional representation, which has an overall skewness coefficient of $\gamma(\alpha_{mean})$:

\begin{definition}[Mean value triangle number]
\begin{multline}\label{EquationMeanTriangleTraditional}
tr^{*}_{mean}(\xi) = \\
$ $\\
\left(m - \,\cos\left(\gamma\left(\alpha_{mean} \pm\,\frac{\pi}{4}\right)\right)\cdot\mathbf{r}(\alpha_{mean}),\: m,\:m - \,\sin\left(\gamma\left(\alpha_{mean} \pm\,\frac{\pi}{4}\right)\right)\cdot\mathbf{r}(\alpha_{mean})\right).
\end{multline}
\end{definition} \sp

This triangle number is uniquely determined and may be referred to as the \emph{mean value triangle} of the fuzzy number $\xi.$

\bigskip
We give a numerical instance returning to \textbf{example~\ref{Example3NonLinear}}: From the definition of skewness~\eqref{EquationSkewnessFuzzyNumberOverall} by the mean value theorem~\eqref{EquationMeanAlphaValue} we receive

\begin{equation}
\alpha_{mean} = 0.6347392094,
\end{equation}
hence
\begin{equation}
\mathbf{r}(\alpha_{mean}) = 3.346605882 , \quad \gamma(\alpha_{mean}) = 0.5025534021,
\end{equation}\sp

\noindent giving a right-skewed fuzzy triangle number $tr^*(l,m,r)$ in traditional notation by
\begin{equation}
\begin{cases}
m-l &= r_{m}\cos\gamma\left(\alpha_{m} + \frac{\pi}{4}\right) = \,0.9339992140,\\
r-m &= r_{m}\sin\gamma\left(\alpha_{m} + \frac{\pi}{4}\right) = \,3.213629785.
\end{cases}
\end{equation}
thus
\begin{equation}
tr^*_{mean}(\xi)(l,m,r) = tr^*(\pi-0.9339992140, \pi, \pi+3.213629785),
\end{equation}\sp

\noindent as depicted below in Fig.~\ref{FigureExample3MeanValueTriangle}.

\begin{figure}[H]
  \centering
  \includegraphics[width=12cm]{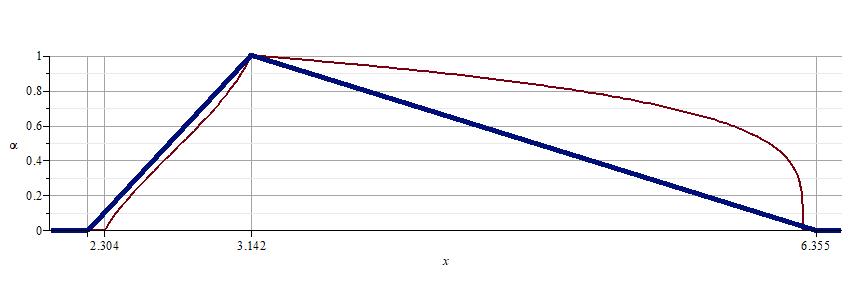}
  \caption{The Mean Value Triangle Number (blue) of a general fuzzy number (dark red).}\label{FigureExample3MeanValueTriangle}
\end{figure}

\begin{remark}
Depending on context it may make more sense to build a triangle of skewness coefficient $\gamma_0$ and magnitude $\mathbf{r}$ {anchored} not in the original fuzzy numbers middle point, but rather in its possibilistic mean value, or in fact any of the single mean values mentioned above, listed for instance in the very readable, comprehensive and accurate study~\cite{BM13}.
\end{remark}

\section{Dispersion as Hausdorff distance from mean value triangle}

Having defined a mean value triangle it suggests itself to take a look at the distance of the fuzzy number to its mean. That is take some form of absolute deviation from it.\sp

It lies in the nature of fuzzy numbers, being a generalization of exact, real intervals that the Hausdorff distance has proven most adequate for many purposes. \sp

It should be noted, that many other metrics have been and constantly are being developed, some of which generalize the Haudorff metric, at many times based on $L^p$ metrics on $[0,1]\times[0,1]$ and mixing both, seldom entirely different approaches. See \cite{TG-RCG09} as a starting point for examples.

Now we may use definition equation~\eqref{EquationMeanTriangleTraditional} to establish a measure of dispersion as the Hausdorff distance of the fuzzy number in question from its own mean value triangle:

\noindent Denote the mean value triangle of a fuzzy number $\xi$ in parametric representation by
\begin{equation}\label{EquationMeanValueTriangleParametric}
tr^*_{mean}(\xi)(\alpha) = [\underline{mean}_{\xi}(\alpha),\overline{mean}_{\xi}(\alpha)].
\end{equation}

As with the skewness coefficient one may alternatively define this dispersion measure level-wise (as a function of $\alpha$) or overall by integrating over $[0,1]$. \sp

\begin{definition}{Dispersion on a given $\alpha$-level}
\begin{equation}\label{EquationDispersionDefinition}
d_{H,\alpha}(\xi) = max\left(|u(\alpha)- \overline{mean}(\alpha)|,|d(\alpha) - \underline{mean}(\alpha)|\right).
\end{equation}
\end{definition}

\noindent In the running example~\ref{Example3NonLinear} we get:
\begin{multline}
d_{H,\alpha}(\xi) =\\
 \max(-3.141592654\alpha^4+1.263726601+1.877866053\alpha,\\
  \abs(-.6862332242+\cos(\alpha+1/3)^2+.6308965082\alpha))
\end{multline}

\noindent which is nicely illustrated in \textit{Fig.~\ref{FigureExample3Dispersion}}:

\begin{figure}[H]
  \centering
  \includegraphics[width=6cm]{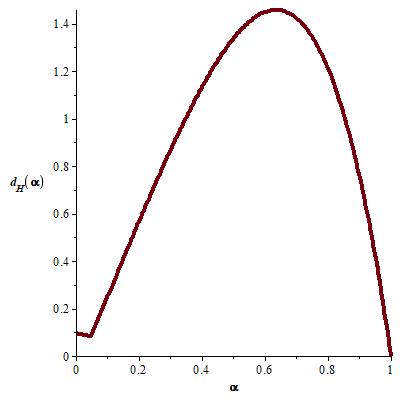}
  \caption{Dispersion: Hausdorff distance of the example fuzzy number $\xi$~\eqref{Example3NonLinear} from its mean value triangle. The distance is easily seen from \textit{Fig.~\ref{FigureExample3MeanValueTriangle}} to determined by the right sides of the FN, down until the lowest (nearing $0$), when the maximum turns the switch. $\alpha$-levels}\label{FigureExample3Dispersion}
\end{figure}

The \emph{overall} dispersion may be defined by integration:\sp

\begin{definition}{Overall dispersion of a fuzzy number.}
\begin{equation}
d_{H}(\xi) = \int_{0}^{1}d_{H,\alpha}\,d\alpha.
\end{equation}
\end{definition}

In the running \textbf{example~\ref{Example3NonLinear}} the overall dispersion of $\xi$ is given by:
\begin{equation}
d_{H}(\xi) = 1.574341097.
\end{equation}

\begin{remark}
The dispersion measure introduced here in~\eqref{EquationDispersionDefinition} is the most obvious choice based on the definition of mean value triangle which again follows naturally from the introduction of our measure of skewness. There is no claim to superiority over any other measures of dispersion being stated in the literature and currently in use. Also countless variations of~\eqref{EquationDispersionDefinition} are possible (by changing the mean value, by changing the distance function), depending on context of the problem at hand, which may suggest a different measure. \sp
The main contribution of this paper remains the new perspective of fuzzy numbers as parameterized curves, and the skewness measure which follows so naturally, without effort, from it.
\end{remark}

\begin{remark}
We use the term \emph{dispersion} to emphasize (set apart not only in essence but also in nomenclature) the difference of fuzzy set theory from probability and statistics, which speaks of most often of variance and standard deviation.
Also is the preference of variance (standard deviation) over average absolute deviation dictated by the algebraic handlebility which it brings. In the case of fuzzy numbers this advantage is not necessary, as unlike is the case with the intricate algebra of random variables (see~\cite{Springer79, Nelson81}) the arithmetic of fuzzy numbers is comparatively clear and simple.
\end{remark}

\section{Fuzzy skewness in Portfolio Optimization Theory}

In this section we make a general case for the use of fuzzy numbers in certain modelling scenarios, when the model inherent uncertainty does not allow for concrete underlying probability distributions to be inferred.

Not wanting to venture into technicalities we begin with a simplified definition outline of the problem:\sp

\subsection{Portfolio selection}
Given a number $I$ of financial assets $\mathbf{i}$, each of which has attributes such as: expected return $r_i$ and volatility $v_i$.\sp

A \emph{portfolio} is a convex choice of weights $w_i \in [0,1]$ to be ascribed to each of those assets, that is a non-negative vector

$$\mathbf{w} = (w_1,w_2, . . . ,w_i), \quad \sum\limits_{i} w_i = 1$$

Portfolio optimization is the decision process of selecting the best portfolio with respect to given objectives $\{\mathbf{obj}_m\}$ and subject to various constraints $\{\mathbf{c}_n\}$. \sp

The primary objective would typically be the portfolio's expected overall return, with secondary objectives including minimization of risk and volatility.\sp

Constraints may relate to budget, sectors, geography, etc. and may be relative with respect to each other
$$\mathbf{c}_{i}= R_{n}(w_{1},..w_{N}) = 0 \quad \text {for instance } \quad w_k \leq w_l,$$
or of absolute nature, e. g.
$$\underline{d}_i \leq \mathbf{i} \leq \overline{d}_i.$$ \sp

Objectives and constraints may be formulated without reference to any given model of uncertainty: probability, fuzziness or other: ``expected return'', `` riskiness'' may be put to numbers by gut feeling, inside knowledge, regression based on historical performance. \sp

But in modern practice all the factors contributing to both the objective functions and constraints are modeled as random variables whose probability distributions in turn are generated / estimated by various methods.\sp

Understanding each asset's return $r_i$ as a random variable, and having determined probability distributions $f_{r_i}$ for each of the assets' return $r_i$ a decision maker has at his disposal the moments $\mu^j_i$ of the random variable, that is $\mu_i^1$ - expected values, $\mu_i^2$ - variance and co-variance, but also skewness, kurtosis and higher moments (See~\cite{HLLP10} for higher moments in portfolio selection).\sp

Set $\mathbf{z} = (r_1,\ldots,r_I),\,$ $\mu_i = E(r_i),\, \mathbf{m} = (\mu_1, \mu_2, . . . , \mu_I)\,$ and $cov({\mathbf{z}}) = \mathlarger{\sum}.$ \sp
If $\mathbf{w} = (w_1,w_2, . . . ,w_I)$ is a set of weights associated with a portfolio, then
the rate of return of this portfolio $$\sum_{i=1}^{I}r_i\cdot w_i$$ is also a random variable with mean $\langle \mathbf{m},\,\mathbf{w} \rangle$ and variance $\mathbf{w}\mathlarger{\sum }\mathbf{w}^{T}.$ \sp

Before Markowitz~\cite{Markowitz52} a portfolio optimization program would look like just

\begin{align}\label{EquationPortfolioSimple}
\text{Maximize}\quad & obj: \sum w_i\cdot \mu_i, \\
\text{subject to}\quad & \quad \{\mathbf{c}_i\},
\end{align}
$\{\mathbf{c_i}\}$ being a set of constraints.

But with the arrival of MPT\footnote{Modern Portfolio Theory} this objective function would change, and an optimal set of weights became one in which the portfolio achieves an acceptable baseline expected rate of return with minimal volatility.
In this theory the variance of the rate of return of an asset is taken as a surrogate for its volatility.\sp

If $\mu_b$ is the aforementioned acceptable baseline expected rate of return, then in the Markowitz theory an optimal portfolio is any portfolio solving the following quadratic program:
\begin{align}\label{EquationPortfolioMatkowitz}
\text{Minimize}\quad & obj: \frac{1}{2}\mathbf{w}\sum \mathbf{w}^{T} \\
\text{subject to}\quad & contstr: \langle \mathbf{m},\,\mathbf{w} \rangle \geq \mu_b, \\
& \text{ and } \{\mathbf{c}_i\}.
\end{align}
where $\{\mathbf{c}_i\},\,$ $\mu_i = E(r_i),\, \mathbf{m} = (\mu_1, \mu_2, . . . , \mu_n),$ and $cov(\mathbf{z}) = \mathlarger{\mathbf{\sum}},$ as above. \sp

\subsection{The mean-variance-skewness model}
With authors such as~\cite{Kane82}, \cite{Lai91} and \cite{LWQ03}, or \cite{VB12, VB13}, \cite{LGY15} for a fuzzy extension, the model has been further refined by bringing in an additional skewness constraint.
- Positive skewness is desirable, since increasing skewness decreases the probability of large negative rates of return. So by bringing in $s_i$ as the skewness of $r_i$ defined by the third moment, and setting a base level of skewness $s_b$ and minimum co-variance $v_b$  one then considers three variations of portfolio optimization programs:

\begin{align}\label{EquationPortfolioMarkowitz}
\text{Minimize}\quad & \frac{1}{2}\,\mathbf{w}^{T}\mathlarger{\sum} \mathbf{w} \\
\text{subject to}\quad & \langle \mathbf{m},\,\mathbf{w} \rangle \geq \mu_b, \\ \nonumber
& \langle \mathbf{s},\,\mathbf{w} \rangle \geq s_b, \\ \nonumber
& \{\mathbf{c}_i\}.
\end{align}

and alternatively

\begin{align}\label{EquationPortfolioMeanMax}
\text{Maximize}\quad & \langle \mathbf{m},\,\mathbf{w} \rangle \\
\text{subject to}\quad & \frac{1}{2}\,\mathbf{w}^{T}\mathlarger{\sum} \mathbf{w}\leq v_b, \\ \nonumber
& \langle \mathbf{s},\,\mathbf{w} \rangle \geq s_b, \\ \nonumber
& \{\mathbf{c}_i\}.
\end{align}

or

\begin{align}\label{EquationPortfolioSkewnessMax}
\text{Maximize}\quad & \langle \mathbf{s},\,\mathbf{w} \rangle \\
\text{subject to}\quad & \langle \mathbf{m},\,\mathbf{w} \rangle \geq \mu_b,\\ \nonumber
& \frac{1}{2}\,\mathbf{w}^{T}\mathlarger{\sum} \mathbf{w}\leq v_b, \\ \nonumber
& \{\mathbf{c}_i\}.
\end{align}

\subsection{Fuzzy Portfolio Optimization}

All of above linear (quadratic) optimization programs may be stated analogically by putting fuzzy numbers in place of random variables on each occurrence.

We give justification to model the uncertainty associated with the assets not by means of probability theory, but resorting to fuzzy set theory instead, and refer to existing literature for existing solution algorithms. \sp

The main, true, core problem in the probabilistic modeling process and its practical implementation is to find the ``right'' probability distributions. \par
When a viable probability distribution can not be found modelers often turn to so-called no-knowledge distributions, such as the uniform, triangular or PERT, setting pessimistic, optimal, optimistic values for each $\mu_i.$  But it must be noted that the underlying assumptions (of \underline{exactly} equal or otherwise placed probabilities) made here are actually very strong, not at all ``no knowledge'', and a fuzzy model, taking interval numbers or fuzzy triangle numbers instead of probabilistic uniform or triangular distributions must be favored. \sp

If a probability distribution governing an investigated process can not be found, it still may often be possible, by various methods, to give sharp lower and upper bounds on the descriptive parameters $\mu_i, cov_{ij}, s_i.$ \sp

In this case, that is with parameters given as interval estimates
\begin{align}
\mu_i = \quad &[\underline{\mu_i},\,\overline{\mu_i}], \\
cov_{ij} = \quad &[\underline{cov_{ij}},\,\overline{cov_{ij}}], \\
s_i = \quad &[\underline{s_i},\,\overline{s_i}],
\end{align}

programs \eqref{EquationPortfolioSimple}, \eqref{EquationPortfolioMarkowitz}, \eqref{EquationPortfolioMeanMax} or \eqref{EquationPortfolioSkewnessMax} fall into the realm of interval linear (quadratic) programming and become tractable as such and their solutions consist of vectors of intervals which may be given explicitly. \sp

For an overview of existing literature on interval programming and leading to more recent results see Milan Hladik's~\cite{Hla09} and~\cite{Hla12}. \sp

Now in consequence of the use of different methods, by different sources, a number of different interval estimates of the investigated parameters may be given. \par
If a number of different interval estimates $\mathbf{E}_n$ are given from a number of sources (experts) these individual estimates may be
aggregated into a single staircase fuzzy number by the procedure given in~\cite{WMGAH15} for type-2 fuzzy intervals:

\begin{equation}\label{MethodByCharacteristicFunction}
\xi(x) = \sum_1^n 1/n\cdot\chi_{E_n}(x).
\end{equation}
with $\chi_{\mathbf{E}_n}$ being the indicator (characteristic) function of $\mathbf{E}_n$.
\begin{equation*}
\chi_{E_n}(x) =
\begin{cases}
1 \quad\text{ if }\quad x\in \mathbf{E_n}, \\
0 \quad\text{ if }\quad x\notin \mathbf{E_n},
\end{cases}
\end{equation*}

\noindent This very intuitive procedure is shown graphically below in figure~\ref{FigureIndividualIntoFuzzy}:

\begin{figure}[H]
\begin{subfigure}{.50\textwidth}
  \centering
  \includegraphics[width=0.9\linewidth]{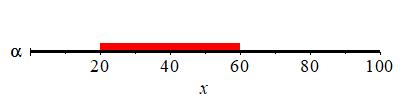}
  \caption{$\mathbf{E}_1$}
  \includegraphics[width=0.9\linewidth]{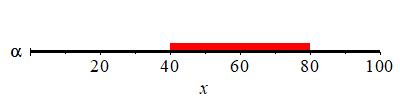}
  \caption{$\mathbf{E}_2$}
  \includegraphics[width=0.9\linewidth]{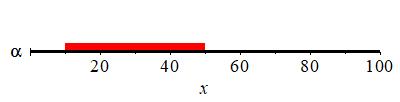}
  \caption{$\mathbf{E}_3$}
\end{subfigure}
\begin{subfigure}{.50\textwidth}
\centering
  \includegraphics[width=0.9\linewidth]{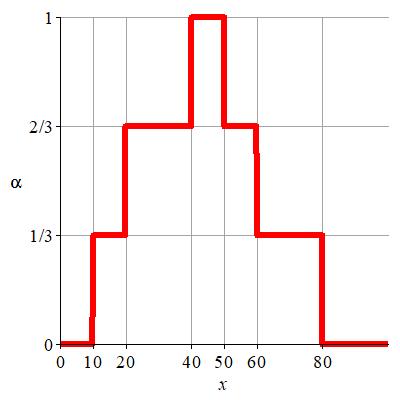}
  \caption{$\xi(x)$}
\end{subfigure}
  \caption{Three aggregated individual interval estimates aggregated into a single staircase fuzzy interval.}\label{FigureIndividualIntoFuzzy}
\end{figure}

\begin{remark}
The method presupposes that there be an overlap of all experts' interval estimates. Because it may be presumed that knowledge and prior information of all experts be sufficiently similar this assumption is quite natural. \sp
In case there is no common overlap of all experts' estimates (the level set at height $\alpha = 1$ is the empty set $\emptyset$), various methods can be applied.
\end{remark}

To facilitate the implementation of the parametric methods discussed in this paper two steps must be taken:

\begin{enumerate}
\item Add another single value level $C_1(\xi)$ on top of the $\alpha$-levels generated by the procedure described in \eqref{MethodByCharacteristicFunction}. This single value can be chosen as the possibilistic mean value or any ranking index which places in the intersection of all constituting $\mathbf{E}_i.$
\item Achieve piecewise differentiability by linearly joining the endpoints at each level (the adaptation of the methods of the preceding sections to piecewise differentiability is routine), or use a mollifier to achieve $C^{\infty},$ or do anything between these two extremes.
\end{enumerate}

The resulting parameterized curve is symbolically graphed below in Fig.~\ref{st}:

\begin{figure}[H]
\begin{subfigure}{.50\textwidth}
  \centering
  \includegraphics[width=0.9\linewidth]{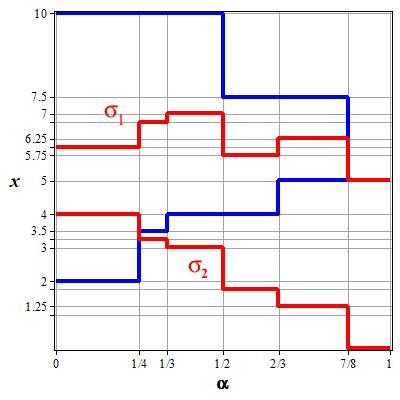}
  \caption{$\xi\,\, \text{and} \,\, F(\xi)$}\label{a}
\end{subfigure}
\begin{subfigure}{.50\textwidth}
\centering
  \includegraphics[width=0.9\linewidth]{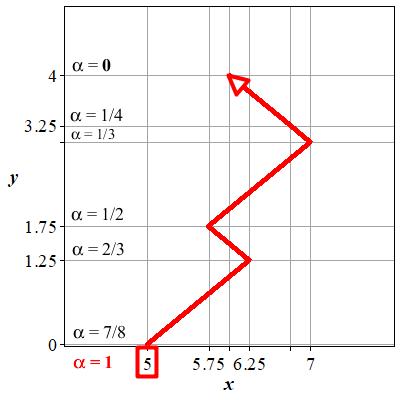}
  \caption{$\sigma(\alpha)$}\label{b}
\end{subfigure}
\caption{A fuzzy staircase number in parametric representation (blue), the sides transformed by $\mathbf{F},$ \eqref{a} and the resulting parameterized curve (on the right, \eqref{b}).}\label{st}
\end{figure}

An much more technical and more sophisticated approach is the attempt to generate parameterized interval estimates from the outset, i.e. upper and lower bounds as $\mu(\alpha) = \left[\underline{\mu}(\alpha),\overline{\mu}(\alpha)\right]$, $v(\alpha) = \left[\underline{v}(\alpha),\overline{v}(\alpha)\right]$ and $s(\alpha) = \left[\underline{s}(\alpha),\overline{s}(\alpha)\right]$, which allows for the construction of fuzzy intervals straightforwardly. This so done in~\cite{SaulTrojani19}. \sp

\bigskip
We consciously refrain from displaying a random generated numerical example of above techniques.

\section{Conclusion}
The main result of this paper is really, that it adds to the here so-called traditional \eqref{EquationOrderedPairDefinition} and parametric \eqref{EquationParametricDefinition} representations of a fuzzy number $\xi$ two other: as a parameterized curve $\sigma(\alpha)$~\eqref{RepresentationCurve}  and as a tangent bundle \eqref{RepresentationTangent}. Then the representation of tangent vectors in polar coordinates $\left(\mathbf{r}(\alpha),\,\gamma(\alpha)\right)$ = (\ref{RepresentationPseudoPolar},\ref{EquationGammaGeneral}) directly implicates measures of skewness at a point~\eqref{EquationSkewnessFuzzyNumberPoint} and overall~\eqref{EquationSkewnessFuzzyNumberOverall}.\sp
The definition of a mean value triangle $tr^*_{\xi}$ \eqref{EquationMeanTriangleTraditional} and dispersion $d_{H}(\xi)$ \eqref{EquationDispersionDefinition} then follow just by going through the motions. \sp

The results achieved in this paper may be developed and diversified in various directions:
\begin{itemize}

\item Comparative studies. \sp
To compare the descriptive parameters, mean value triangle, dispersion, and skewness, developed in this paper to measures which have been developed before.

\item Go into Sobolev spaces, \sp
 to include non-differentiable membership functions. As stated in~\textbf{remark~\ref{RemarkSobolev}}:
Much less restrictive assumptions may be adopted and still achieve all results of this paper.

\item Parametric curves as primary representation: \sp
The arrow/ vector representation of fuzzy triangle numbers may appear more intuitive to non-mathematical decision makers and takers than the traditional and parametric ones.

\item Differential geometry of curves: \sp

This paper hints at, but does not exploit curvature, torsion, generally the \textbf{TNB} frame of a fuzzy number understood as a parameterized curve.

\item Computer aided visualisation and eye-tracking: \sp

Visualizing the twists and turns of a linguistic variable.

\end{itemize}

\bibliographystyle{elsarticle-num}

\begin{thebibliography}{10}
\expandafter\ifx\csname url\endcsname\relax
  \def\url#1{\texttt{#1}}\fi
\expandafter\ifx\csname urlprefix\endcsname\relax\def\urlprefix{URL }\fi
\expandafter\ifx\csname href\endcsname\relax
  \def\href#1#2{#2} \def\path#1{#1}\fi

\bibitem{Pearson1895}
K.~Pearson, Philosophical Transactions of the Royal Society of London.
  A\href{http://www.jstor.org/stable/90649}{[link]}.
\newline\urlprefix\url{http://www.jstor.org/stable/90649}

\bibitem{Zwe64}
W.~R. von Zwet, Convex transformations: A new approach to skewness and
  kurtosis, Springer New York, New York, NY, 2012, pp. 3--11.

\bibitem{Men51}
K.~Menger, Ensembles flous et fonctions al\'{e}atoires., Comptes rendus
  hebdomadaires des seances de l'academie des sciences 232~(22) (1951)
  2001--2003.

\bibitem{GM84}
R.~A. Groeneveld, G.~Meeden,
  \href{http://www.jstor.org/stable/2987742}{Measuring skewness and kurtosis},
  Journal of the Royal Statistical Society. Series D (The Statistician) 33~(4)
  (1984) 391--399.
\newline\urlprefix\url{http://www.jstor.org/stable/2987742}

\bibitem{GV86}
R.~Goetschel~Jr., W.~Voxman, Elementary fuzzy calculus, Fuzzy Sets and Systems
  18~(1) (1986) 31--43.
\newblock \href {https://doi.org/10.1016/0165-0114(86)90026-6}
  {\path{doi:10.1016/0165-0114(86)90026-6}}.

\bibitem{CF01}
C.~Carlsson, R.~Full\'{e}r, On possibilistic mean value and variance of fuzzy
  numbers, Fuzzy Sets and Systems 122~(2) (2001) 315--326.
\newblock \href {https://doi.org/10.1016/S0165-0114(00)00043-9}
  {\path{doi:10.1016/S0165-0114(00)00043-9}}.

\bibitem{FM03}
R.~Full\'{e}r, P.~Majlender, On weighted possibilistic mean and variance of
  fuzzy numbers, Fuzzy Sets and Systems 136~(3) (2003) 363--374.
\newblock \href {https://doi.org/10.1016/S0165-0114(02)00216-6}
  {\path{doi:10.1016/S0165-0114(02)00216-6}}.

\bibitem{VB12}
E.~Vercher, J.~Berm\'{u}dez, Fuzzy portfolio selection models: A numerical
  study, Springer Optimization and Its Applications 70 (2012) 253--280.

\bibitem{VB13}
E.~Vercher, J.~Berm\'{u}dez, A possibilistic mean-downside risk-skewness model
  for efficient portfolio selection, IEEE Transactions on Fuzzy Systems 21~(3)
  (2013) 585--595.

\bibitem{LGY15}
X.~{Li}, S.~{Guo}, L.~{Yu}, Skewness of fuzzy numbers and its applications in
  portfolio selection, IEEE Transactions on Fuzzy Systems 23~(6) (2015)
  2135--2143.
\newblock \href {https://doi.org/10.1109/TFUZZ.2015.2404340}
  {\path{doi:10.1109/TFUZZ.2015.2404340}}.

\bibitem{Buck02}
J.~Buckley, E.~Eslami, An Introduction to Fuzzy Sets and Fuzzy Logic,
  Physica-Verlag Heidelberg New York (A Springer-Verlag Company), 2002.

\bibitem{Vie11}
R.~Viertl, Statistical Methods for Fuzzy Data, John Wiley $\&$ Sons, 2011.

\bibitem{Zad75.1}
L.~Zadeh, The concept of a linguistic variable and its application to
  approximate reasoning-i, Information Sciences 8~(3) (1975) 199--249.
\newblock \href {https://doi.org/10.1016/0020-0255(75)90036-5}
  {\path{doi:10.1016/0020-0255(75)90036-5}}.

\bibitem{Zad75.2}
L.~Zadeh, The concept of a linguistic variable and its application to
  approximate reasoning-ii, Information Sciences 8~(4) (1975) 301--357.
\newblock \href {https://doi.org/10.1016/0020-0255(75)90046-8}
  {\path{doi:10.1016/0020-0255(75)90046-8}}.

\bibitem{Zad75.3}
L.~Zadeh, The concept of a linguistic variable and its application to
  approximate reasoning-iii, Information Sciences 9~(1) (1975) 43--80.
\newblock \href {https://doi.org/10.1016/0020-0255(75)90017-1}
  {\path{doi:10.1016/0020-0255(75)90017-1}}.

\bibitem{Zad97}
L.~Zadeh, Toward a theory of fuzzy information granulation and its centrality
  in human reasoning and fuzzy logic, Fuzzy Sets and Systems 90~(2) (1997)
  111--127.

\bibitem{BKKN97}
B.~Bouchon-Meunier, O.~Kosheleva, V.~Kreinovich, H.~Nguyen, Fuzzy numbers are
  the only fuzzy sets that keep invertible operations invertible, Fuzzy Sets
  and Systems 91~(2) (1997) 155--163.
\newblock \href {https://doi.org/10.1016/S0165-0114(97)00137-1}
  {\path{doi:10.1016/S0165-0114(97)00137-1}}.

\bibitem{BM13}
M.~Brunelli, J.~Mezei, How different are ranking methods for fuzzy numbers? a
  numerical study, International Journal of Approximate Reasoning 54~(5) (2013)
  627--639.
\newblock \href {https://doi.org/10.1016/j.ijar.2013.01.009}
  {\path{doi:10.1016/j.ijar.2013.01.009}}.

\bibitem{DP87}
D.~Dubois, H.~Prade, The mean value of a fuzzy number, Fuzzy Sets and Systems
  24~(3) (1987) 279 -- 300, fuzzy Numbers.
\newblock \href {https://doi.org/https://doi.org/10.1016/0165-0114(87)90028-5}
  {\path{doi:https://doi.org/10.1016/0165-0114(87)90028-5}}.

\bibitem{Dem67}
A.~P. Dempster, \href{https://doi.org/10.1214/aoms/1177698950}{Upper and lower
  probabilities induced by a multivalued mapping}, Ann. Math. Statist. 38~(2)
  (1967) 325--339.
\newblock \href {https://doi.org/10.1214/aoms/1177698950}
  {\path{doi:10.1214/aoms/1177698950}}.
\newline\urlprefix\url{https://doi.org/10.1214/aoms/1177698950}

\bibitem{DVW98}
M.~Delgado, M.~Vila, W.~Voxman, On a canonical representation of fuzzy numbers,
  Fuzzy Sets and Systems 93~(1) (1998) 125--135.
\newblock \href {https://doi.org/10.1016/S0165-0114(96)00144-3}
  {\path{doi:10.1016/S0165-0114(96)00144-3}}.

\bibitem{DVW298}
M.~Delgado, M.~Vila, W.~Voxman, A fuzziness measure for fuzzy numbers:
  Applications, Fuzzy Sets and Systems 94~(2) (1998) 205--216.
\newblock \href {https://doi.org/10.1016/S0165-0114(96)00247-3}
  {\path{doi:10.1016/S0165-0114(96)00247-3}}.

\bibitem{DK99}
P.~Diamond, P.~Kloeden, Metric spaces of fuzzy sets, Fuzzy Sets and Systems
  100~(SUPPL. 1) (1999) 63--71.

\bibitem{TG-RCG09}
W.~Trutschnig, G.~Gonz\'{a}lez-Rodr\'{\i}guez, A.~Colubi, M.~Gil, A new family
  of metrics for compact, convex (fuzzy) sets based on a generalized concept of
  mid and spread, Information Sciences 179~(23) (2009) 3964--3972.
\newblock \href {https://doi.org/10.1016/j.ins.2009.06.023}
  {\path{doi:10.1016/j.ins.2009.06.023}}.

\bibitem{Springer79}
M.~Springer, The Algebra of Random Variables, Probability and Statistics
  Series, Wiley, 1979.

\bibitem{Nelson81}
P.~R. Nelson, The algebra of random variables, Technometrics 23~(2) (1981)
  197--198.
\newblock \href {https://doi.org/10.1080/00401706.1981.10486266}
  {\path{doi:10.1080/00401706.1981.10486266}}.

\bibitem{HLLP10}
C.~Harvey, J.~Liechty, M.~Liechty, M.~Peter, Portfolio selection with higher
  moments, Quantitative Finance 10~(5) (2010) 469--485.
\newblock \href {https://doi.org/10.1080/14697681003756877}
  {\path{doi:10.1080/14697681003756877}}.

\bibitem{Markowitz52}
H.~Markowitz, Portfolio selection*, The Journal of Finance 7~(1) (1952) 77--91.
\newblock \href {https://doi.org/10.1111/j.1540-6261.1952.tb01525.x}
  {\path{doi:10.1111/j.1540-6261.1952.tb01525.x}}.

\bibitem{Kane82}
A.~Kane, \href{http://www.jstor.org/stable/2330926}{Skewness preference and
  portfolio choice}, The Journal of Financial and Quantitative Analysis 17~(1)
  (1982) 15--25.
\newline\urlprefix\url{http://www.jstor.org/stable/2330926}

\bibitem{Lai91}
T.-Y. Lai, Portfolio selection with skewness: A multiple-objective approach,
  Review of Quantitative Finance and Accounting 1~(3) (1991) 293--305.
\newblock \href {https://doi.org/10.1007/BF02408382}
  {\path{doi:10.1007/BF02408382}}.

\bibitem{LWQ03}
S.~Liu, S.~Y. Wang, W.~Qiu, Mean-variance-skewness model for portfolio
  selection with transaction costs, International Journal of Systems Science
  34~(4) (2003) 255--262.
\newblock \href {https://doi.org/10.1080/0020772031000158492}
  {\path{doi:10.1080/0020772031000158492}}.

\bibitem{Hla09}
M.~Hladík, Optimal value range in interval linear programming, Fuzzy
  Optimization and Decision Making 8~(3) (2009) 283--294.
\newblock \href {https://doi.org/10.1007/s10700-009-9060-7}
  {\path{doi:10.1007/s10700-009-9060-7}}.

\bibitem{Hla12}
M.~Hlad\'{\i}k, Interval linear programming: A survey, 2012.

\bibitem{WMGAH15}
C.~Wagner, S.~Miller, J.~Garibaldi, D.~Anderson, T.~Havens, From
  interval-valued data to general type-2 fuzzy sets, IEEE Transactions on Fuzzy
  Systems 23~(2) (2015) 248--269.
\newblock \href {https://doi.org/10.1109/TFUZZ.2014.2310734}
  {\path{doi:10.1109/TFUZZ.2014.2310734}}.

\bibitem{SaulTrojani19}
P.~Schneider, F.~Trojani, (almost) model-free recovery, The Journal of Finance
  74~(1) (2019) 323--370.
\newblock \href {https://doi.org/10.1111/jofi.12737}
  {\path{doi:10.1111/jofi.12737}}.

\end{thebibliography}

\end{document}